\documentclass[a4paper]{article}
\usepackage{a4wide}
\usepackage[normalem]{ulem}
\usepackage{caption}
\usepackage{float}
\usepackage{subfig}
\usepackage{dsfont}
\usepackage[T1]{fontenc}
\usepackage{amsmath}
\usepackage{amssymb}
\usepackage{amsthm}
\usepackage[ansinew]{inputenc}
\usepackage{enumerate}
\usepackage{fancyhdr}
\usepackage{color}
\usepackage{ifpdf}
\usepackage{graphicx}
\usepackage{rotating}
\usepackage[T2A]{fontenc}
\usepackage{mathtext}
\usepackage{caption}
\theoremstyle{plain}

\newtheorem*{thm*}{Theorem}

\newtheorem*{prop*}{Proposition}

\usepackage{authblk}

\begin{document}

\title{A Dynamical Systems Analysis of Axisymmetric Accretion in a Schwarzschild Black Hole}
\author[1]{Shamreen Iram\footnote{samreen654@gmail.com}}
\affil[1]{Indian Institute of Technology, Kanpur, Uttar Pradesh-208016, India}
\maketitle

\begin{abstract}
Stationary, inviscid, axisymmetric, rotating, transonic accretion flow has been studied in a general relativistic framework, in the Schwarzschild metric; for three different flow geometries - under both polytropic and isothermal conditions. The equilibrium points of the underlying fluid system have been located and an eigenvalue based linear dynamical systems analysis of these critical points has been carried out, to obtain a taxonomic scheme of the critical points. It has hence been shown that only saddle and centre type points can arise for real, physical transonic flow.  
\end{abstract}

\section{Introduction}

Current research on alternative models to gravity has been spurred by the search for possible systems that are more feasible 'study' candidates than strong gravity objects like astrophysical black holes- open to exploration and analysis at length scales we can more easily probe. A one to one correspondence between such systems and gravitational systems would then lead to verification of rich and singular properties of exotic strongly gravitating bodies. This was the main motivation behind the rise of analogue gravity theory, where perturbations propagating in a fluid medium are identified as doing so on a relativistic acoustic geometry embedded by the background stationary space time, the erstwhile 'acoustic metric'. A complete mapping may now be established between the physics playing out on this template, and the dynamical features of space-time as perceived in general theory of relativity. The advantage of this formalism is that the regime of postulated Hawking radiation now shifts from the trans-Planckian to the trans-Bohrian zone, which is well within our explorable domain. \cite{Pu12}\\

An astrophysical black hole in a matter rich environment or with access to a matter source, will accrete matter towards itself, leading to the formation of an accretion disc. The accreting source point is assumed to be located far enough from the accretor (i.e black hole) for it's field to be sufficiently weak to allow the matter flow to start out moving very slow, and thus at subsonic velocities.\cite{Goswami,Nag11,Tarafdar} It speeds up along the way as it spirals inwards. The black hole inner boundary condition implies perforce that the flow must be supersonic when it crosses the event horizon.\cite{Beg78,Brink80,Shap83,Ch90} Thus, unlike accretion by other candidate objects of strong gravity, black hole accretion must necessarily be transonic in nature. This indicates that an accreting black hole system can be considered as a classical analogue gravity model naturally found in the universe and such a model demonstrates a unique and exquisite feature - it contains both gravitational as well as acoustic horizons. More than one acoustic horizon are formed for the multi transonic flow.\cite{Abrm81,Fukue87,Ch89,Ch90,Das07} The transition from subsonic to supersonic flow can proceed either smoothly and continuously \cite{Ch90}, or via discontinuous\cite{Ch89,Goswami}, violently dissipative shock transitions\cite{Goswami}.\\

On another note, the importance of accretion studies in black hole physics increases several folds when one realises that in the context of garnering direct observational evidence of them, spectroscopy, the traditional go-to tool for astrophysicists, has not been of much help. The obscuring presence of the event horizon would make any such ’visibility’ nigh impossible. One must then take recourse to the fact that an astrophysical black hole, being a strongly gravitating body, should leave several gravitational signatures upon its surrounding space-time. These could be used as indirect evidence of its presence. Of specific motivational relevance to this study are the radiative emissions that could be caused by infalling matter accreting on to a black hole when rest mass undergoes conversion to radiation. An accreting black hole may be detected this way. Other gravitationally driven phenomena spurring great interest in current research include the possibility of remarkable ’lensing’ effects that could be produced by the deep gravitational potential wells around isolated black holes\cite{Zhang}.\\

This study has been carried out for stationary, inviscid, non-self gravitating, hydrodynamic axi-symmetric accretion in the low angular momentum regime. Three different accretion flow geometries have been studied- namely, constant height disc geometry, conical disc geometry, and a disc in vertical equilibrium. Such slowly rotating spiraling flows are not merely theoretical abstractions, as has been stated in previous literature \cite{Tarafdar,Nag11}. Several astrophysical systems like detached binary systems fed by accretion from 0B stellar winds \cite{Ill75,Li80}, semi-detached low magnetic binaries\cite{Bisi98}, and supermassive black holes fed by accretion from slowly rotating central stellar clusters \cite{Ill88,Ho99} provide prime examples. Of the considered geometric configurations, the constant height model is the simplest possible model, of a disc showing translational symmetry along its axis of rotation. The disc in conical equilibrium shows radial symmetry, and is the closest approximation to a slowly rotating disc. The disc in vertical equilibrium is the most widely considered, and disc thickness is a function of local disc radius and other parameters like sound speed. Thermodynamic conditions pertaining to different equations of state have been considered in this study, varying from a range of polytropic conditions, ending with pure isothermal. Real, physical transonic solutions connecting our zones of interest (the event horizon with the source of accreting material, effectively infinity) may be formally obtained as critical solutions on the phase plane of the flow- spanning flow velocity and radial distance. By virtue of simultaneously satisfying the conditions of physicality and transonicity, these solutions must necessarily be associated with saddle type critical points- the only kind that will allow the solution to double up and pass through itself. Mention must be made at this point that phase space analyses of the underlying fluid equations throws up critical points on the phase trajectory, and for certain zones of low angular momentum accretions, criticality shows up recurrently, giving rise to 'multicritical' zones - a rich feature of axisymmetric black hole accretion. While multicriticality is taken to imply multitransonicity and the two are used interchangeably here, this need not be the case. One possible way for multiple acoustic horizons to form in the system  is to proceed via shock transitions between two such points. Thus shock solutions need to be accessible and the energetic and thermal conditions pertaining to their formation need to be satisfied.\cite{Das07,Das12} Of the three critical point multicritical zone that is seen to arise in this work, stationary shock solutions usually form between the outer and middle critical points\cite{Pu12}. Standing shocks in such systems are of especial significance within the paradigm of accretion, since the violently dissipative processes at play at these shock fronts provide a potential source of radiative emissions, by disrupting the ordered flow of infalling matter. \\

In this paper, the fundamental, governing non-linear equations for stationary, inviscid, hydro-dynamic, rotating axisymmetric accretion flow in the Schwarzschild metric have been mapped onto a first order, autonomous dynamical system \cite{Jordan,Bohr93,Ray02,Afs03,Ch06,Mandal07}. A full and comprehensive global picture of the phase space topology will involve taking recourse to complete numerical integration of the system of non-linear coupled differential flow equations. Here, an alternative quasi-analytical approach has been presented which employs further dynamical systems based techniques to gain further in- sight into the nature of these critical points. First, the critical points of the flow have been located computationally and the parameter space spanned by the solutions identified. Thereafter, the autonomous system has been subsequently linearized about these critical points and a complete, rigorous and equivalent, eigenvalue based classification formalism has been developed to identify the nature of the critical points. This has been carried out for three different prior mentioned geometric flow configurations, and various equations of state. It has hence been shown that the only type of critical points possible on the flow trajectory are of saddle or centre type. The treatment and procedure presented in this paper has been employed in earlier works, and follows and builds up from them. One work has carried out the study for a pseudo-Newtonian potential\cite{Paczynski80,Nag11}. While another has investigated the procedure in the Kerr metric, for the case of polytropic flow in a disc in vertical equilibrium, and examined the result in the Schwarzschild limit\cite{Goswami}. This work extends it to other flow geometries, and all possible equations of state (namely a range of polytopic to isothermal),in the Schwarzschild metric.

\section{Background Configuration of the Flow and Critical Points}

The subject of this treatment is on self-gravitating axis-symmetric, inviscid hydrodynamic accretion onto a non-rotating, i.e., Schwarzschild black hole, in the low angular momentum regime.\\

Such slowly rotating spiraling flows are not merely theoretical abstractions, as has been stated in previous literature \cite{Tarafdar, Nag11}. Several astrophysical systems like detached binary systems fed by accretion from 0B stellar winds \cite{Ill75,Li80}, semi-detached low magnetic binaries\cite{Bisi98}, and supermassive black holes fed by accretion from slowly rotating central stellar clusters \cite{Ill88,Ho99} provide prime examples.\\

In such a background configuration, the fundamental governing equations of the flow have been first set up, under polytropic thermodynamic conditions, followed by accretion under isothermal conditions.

\subsection{Polytropic Accretion}

Bernoulli's equation can be expressed analogously in its relativistic form \cite{And89,Bar04,Das07,Goswami}\\

\begin{equation}
\varepsilon = h v_t
\end{equation}

where $\varepsilon$ is the specific flow energy, $h$ is the specific enthalpy of the system and $v_t$ is the temporal component of the 4-velocity given by\\

\begin{equation}
v_t = \sqrt{\frac{1-2/r}{\bigg[1-\frac{\lambda^2}{r^2}(1-2/r)\bigg](1-u^2)}}
\end{equation}

Here, $\lambda$ is the angular momentum parameter and $ u $ is the advective fluid velocity.\\

The equation of state is taken to be of the form\\

\begin{equation}
P = k \rho^{\gamma}
\end{equation}

where $\gamma$ is the usual polytropic index, the ratio of the specific heats at constant pressure $C_p$ and at constant volume $C_v$. $k$ provides a measure of the specific entropy of the accreting fluid, at constant entropy \cite{Tarafdar} .\\

The specific enthalpy is of the form\\

\begin{equation}
h = \frac{P+\epsilon}{\rho}
\end{equation}

where energy density $\epsilon$ is a function of the rest mass density and internal energy, and is defined as \\

\begin{equation}
\epsilon = \rho + \frac{P}{\gamma - 1 }
\end{equation}

Plugging in the above defined equation in the equation of state, the definition for advective sound speed becomes:\\

\begin{equation}
c_s^2 = \frac{\partial P}{\partial \epsilon}\bigg|_{h=const.}
\end{equation}

Now, at constant entropy, the enthalpy can be expressed as \\

\begin{equation}
h=\frac{\partial \epsilon}{\partial \rho} 
\end{equation}

This defines the expression for $h$ as \\

\begin{equation}
h=\frac{\gamma-1}{\gamma-(1+c_s^2)}
\end{equation}

The final explicit expression for conserved $\varepsilon$ is given as \cite{Tarafdar}

\begin{equation}
\varepsilon = -\frac{\gamma-1}{\gamma-(1+c_s^2)}\sqrt{\frac{1-2/r}{\bigg[1-\frac{\lambda^2}{r^2}(1-2/r)\bigg](1-u^2)}}
\end{equation}

The following transformation of the conserved mass accretion rate $\dot{M}$, given by \\

\begin{equation}
\dot{\Xi} = \dot{M}K^{1/(\gamma-1)}\gamma^{1/(\gamma-1)}
\end{equation}

has been used to locate and separate out the accretion and the wind zones, in the multi-critical parts of the parameter space. Here, K gives a measure of the specific entropy of the infalling accreting matter and $\dot{\Xi}$ can be physically identified as the entropy accretion rate. \\

\subsubsection{Constant Height Model}

The disc height H plays a trivial role in this model, which shows translational symmetry along its vertical axis. The conserved mass accretion rate is \cite{Tarafdar}:

\begin{equation}
\dot{M} = 2 \pi \rho u \sqrt{\frac{1-2/r}{1-u^2}} rH
\end{equation}

which defines 

\begin{equation}
\dot{\Xi} = 2\pi u \sqrt{\frac{1-2/r}{1-u^2}} r c_s^{\frac{2}{\gamma-1}}\bigg[\frac{\gamma-1}{\gamma-(1+c_s^2)}\bigg]^{\frac{1}{\gamma-1}}H
\end{equation}

Differentiating the two conserved quantities $\varepsilon$ and $\dot{\Xi}$ and combining the two obtained relations between $\frac{dc_s}{dr}$ and $\frac{du}{dr}$ gives the following final spatial value of change of advective velocity \cite{Tarafdar} \\

\begin{equation}
\frac{du}{dr} = \frac{c_s^2[\frac{1}{r}+\frac{1}{r^2(1-2/r)}]-f_2(r,\lambda)}{(1-c_s^2)\frac{u}{1-u^2}-\frac{c_s^2}{u}} 
= \frac{N}{D}
\end{equation}

where 

\begin{equation}
f_2 = -\frac{\lambda^2}{r^3}\bigg[\frac{1-3/r}{1-\frac{\lambda^2}{r^2}(1-2/r)}\bigg]+\frac{1}{r^2(1-2/r)}
\end{equation}

The critical point conditions then become fixed on the u-r phase plane as
 (setting N and D to 0)

\begin{equation}
[u = c_s]_{r_c} \hspace{1 cm} 
[c_s]_{r_c} = \sqrt{\frac{f_2}{\frac{1}{r_c}+\frac{1}{r_c^2(1-2/r_c)}}}
\end{equation}

$[\varepsilon, \lambda, \gamma]$ and relation (15) have now been used to numerically locate the critical points of the flow, and demarcate the mono and the multi-linear zones in parameter space. \\

From (15), the critical points are seen to be identical to the sonic points, the point where the advective and sound speeds equalise. \\

The astrophysically relevant domain for the above numerical computations are 
$ [1\leq \varepsilon \leq 2]$,
$[0 \leq \lambda \leq 4]$, $[\frac{4}{3} \leq \gamma \leq\frac{5}{3}]$ \cite{Tarafdar}

\subsubsection{Flow in Conical Equilibrium}

The height is now taken to be quasi-spherical, and mass accretion and entropy accretion rates are appropriately modified to : \cite{Tarafdar}\\

\begin{eqnarray}
\dot{M} &=& \Lambda \rho u \sqrt{\frac{1-2/r}{1-u^2}}r^2\\
\dot{\Xi} &=& \Lambda u\sqrt{\frac{1-2/r}{1-u^2}}r^2 c_s^{\frac{2}{\gamma-1}}\bigg[\frac{\gamma-1}{\gamma-(1+c_s^2)}\bigg]^{\frac{1}{\gamma-1}} 
\end{eqnarray}

where $\Lambda$ is a constant angular factor determining the radial shape of the flow. Proceeding as above,

\begin{equation}
\frac{du}{dr} = -\frac{c_S^2\bigg[\frac{2}{r}+\frac{1}{r^2(1-2/r)}\bigg] - f_2}{\frac{u}{1-u^2}(c_s^2-1)-\frac{c_s^2}{u}}
\end{equation}

giving the critical point conditions :

\begin{eqnarray}
[u = c_s]_{r_c} \hspace{1cm} 
[c_s]_{r_c} = \sqrt{\frac{f_2}{\frac{2}{r_c}+\frac{1}{r_c^2(1-2/r_c)}}}
\end{eqnarray}

Here too, the critical and the sonic points are seen to coincide.\\

\subsubsection{Flow in Vertical Equilibrium}

The disc height for flow in the hydrostatic equilibrium in vertical direction can be taken to be \cite{Das07}

\begin{equation}
H(r) = \frac{r^2 c_s}{\lambda}\sqrt{\frac{2(1-u^2)[1 - \frac{\lambda^2}{r^2}(1 - 2/r)](\gamma -1)}{\gamma(1-2/r)[\gamma - (1+c_s^2)]}}
\end{equation}

Mass and entropy accretion become\cite{Tarafdar}

\begin{eqnarray}
\dot{M} = 4\pi \rho \sqrt{\frac{u(1-2/r)}{1-u^2}} \frac{r^4 c_s}{\lambda}\sqrt{\frac{2(1-u^2)[1 - \frac{\lambda^2}{r^2}(1 - 2/r)](\gamma -1)}{\gamma(1-2/r)[\gamma - (1+s^2)]}}\\
\dot{\Xi} = \sqrt{\frac{2}{\gamma}}\bigg[\frac{\gamma -1}{\gamma-(1+c_s^2)}\bigg]^{\frac{\gamma+1}{2(\gamma-1)}}\frac{c_s^{\frac{\gamma-1}{\gamma+1}}}{\lambda}\sqrt{1-\frac{\lambda^2}{r^2}(1-2/r)}(4 \pi u r^3)
\end{eqnarray}

giving \cite{Tarafdar}

\begin{equation}
\frac{du}{dr} = \frac{\frac{2c_s^2}{\gamma+1}f_1(r,\lambda) - f_2(r,\lambda)}{\frac{u}{1-u^2} - \frac{2c_s^2}{u(\gamma+1)}}
\end{equation}

where $f_1(r,\lambda) = \frac{3}{r}+\frac{\lambda^2}{r^3}\bigg[\frac{1-3/r}{1-\frac{\lambda^2}{r^2}(1-2/r)}\bigg]$

The critical point conditions come out to be

\begin{equation}
u_c = \sqrt{\frac{1}{1+(\frac{\gamma+1}{2})\frac{1}{c_{sc}^2}}} = \sqrt{\frac{f_2}{f_1 + f_2}}
\end{equation}

Therefore, it can be seen that the critical and the sonic points are not isomorphic.

\subsection{Isothermal Accretion}

For isothermal accretion flow, the equation of state considered is of the form

\begin{equation}
p = \rho c_s^2 = \frac{R}{\mu} \rho T = \frac{\rho k_B}{\mu m_H}T
\end{equation}

where $R$, $k_B$, $\mu$, $m_H$ and $T$ are the universal gas constant, the Boltzmann constant, the reduced mass, mass of the hydrogen atom and the isothermal flow temperature. It is evident from above that the isothermal sound speed $c_s$ is thus a position independent global flow constant. Under such condition, the static general relativistic fluid equations deliver the following first integral of motion independent of disc geometry \cite{Tarafdar}

\begin{equation}
\xi = \frac{r^2(r-2)}{[r^3-(r-2)\lambda^2](1-u^2)}\rho^{2c_s^2}
\end{equation}

This is the analogous conserved quantity for isothermal flow corresponding to conserved energy $\varepsilon$ for polytropic accretion.\\

For this case, the sound speed being a constant, $\xi$ has been used to separate out the accretion and wind zones after locating the multi-critical parts of parameter space.

\subsubsection{Constant Height Model}

The mass accretion rate and $\xi$ together give \cite{Tarafdar}

\begin{equation}
\frac{du}{dr} = \frac{[2r^3 - 2(r-2)^2\lambda^2+(1-r)(2r^3+4\lambda^2-2r\lambda^2)c_s^2]u(u^2-1)}{r(2-r)[-2r^3-4\lambda^2+2r\lambda^2](u^2-c_s^2)}
\end{equation}

delivering the critical point conditions as 

\begin{equation}
[u=c_s]_{r_c} \hspace{1 cm}
c_s^2 = \bigg[\frac{-r^3+(r-2)^2\lambda^2}{r^3-r^4+(r-2)(r-1)\lambda^2} \bigg]_{r_c}
\end{equation}

Using the value of $c_s^2$ from the Clayperon-Mendeleev equation as input above, and taking T and $\lambda$ as parameters , the critical points have now been located. Further, from $\dot{M}$ (which is a constant), plugging in $\rho$ back into $\xi$, the accretion and wind zones have been separated out.\\

Like in the polytropic case, the critical and the sonic points for this disc geometry are seen to be degenerate.

\subsubsection{Disc in Conical Equilibrium}

The space gradient of advective velocity comes out to be\cite{Tarafdar}:

\begin{equation}
\frac{du}{dr}=\frac{[2r^3-2(r-2)^2\lambda^2+(3-2r)(2r^3+4\lambda^2-2r\lambda^2)c_s^2]u(u^2-1)}{r(2-r)[-2r^3-4\lambda^2+2r\lambda^2](u^2-c_s^2)}
\end{equation}

The critical point conditions are fixed as

\begin{equation}
[u=c_s]_{r_c}\hspace{1 cm}
c_s^2 = \bigg[\frac{-r^3+(r-2)^2\lambda^2}{3r^3-2r^4+6\lambda^2-7r\lambda^2+2r^2\lambda^2}\bigg]_{r_c}
\end{equation}

\subsubsection{Flow in Vertical Equilibrium}

Unlike the other two models, taking constant sound speed, the disc height for flow in this 
geometric configuration modifies to \cite{Tarafdar}:

\begin{equation}
H(r)^{iso} = rc_s\frac{\sqrt{2[r^3-(r-2)\lambda^2](1-u^2)}}{\lambda\sqrt{r-2}}
\end{equation}

giving mass accretion rate

\begin{equation}
\dot{M} = 4\pi \rho \frac{r^2uc_s}{\lambda}\sqrt{2[r^3-(r-2)\lambda^2]}
\end{equation}

and \cite{Tarafdar}:

\begin{equation}
\frac{du}{dr}=\frac{2[r^3-(r-2)^2\lambda^2+(2-r)(4r^3+5\lambda^2-3r\lambda^2)c_s^2]u(u^2-1)}{r(r-2)[-2r^3-4\lambda^2+2r\lambda^2][c_s^2-(1+c_s^2)u^2]}
\end{equation}

They give the following critical pair conditions\cite{Tarafdar}:

\begin{equation}
\bigg[u^2=\frac{c_s^2}{1+c_s^2} = \frac{-r^3+(r-2)^2\lambda^2}{8r^3-4r^4+10\lambda^2-11r\lambda^2+3r^2\lambda^2}\bigg]_{r_c}
\end{equation}

As is evident, unlike the cases of constant height and disc in conical equilibrium, the sonic points and critical points do not coincide in this case and are not isomorphic. Thus for general relativistic isothermal flow in the Schwarzschild metric, the Mach number at the critical surface is not 1. It is interesting to note at this point that for isothermal axis-symmetric accretion in Newtonian \cite{Ch06} and pseudo-Schwarzschild \cite{Nag11} potentials, however, for the case of the disc in vertical equilibrium too, the sonic points had coincided with the critical points. This feature did not carry over into the general relativistic regime.

\section{Nature of the Fixed Points : A Dynamical Systems Study}

To carry out a dynamical systems study of the fixed points of the flow, the equations for space gradient of advective velocity for the stationary, axisymmetric rotational flow in the Schwarzschild metric have been first appropriately parametrised and decomposed into an equivalent autonomous first-order dynamical system, in the independent parameter $\tau$ \cite{Jordan,Bohr93,Ray02,Afs03,Ch06,Mandal07}. \\

It must be noted here that the equations have now been taken in the $u^2 - r$ plane, to remove any directionality dependence.\\

\subsection{Polytropic Accretion}

The following linear perturbation scheme has been applied about the critical point : 

\begin{eqnarray*}
u^2 &=& u_c^2 +\delta u^2\\
c_s^2 &=& c_{sc}^2 + \delta c_s^2\\
r &=& r_c +\delta r
\end{eqnarray*}

Now, the conserved energy equation for polytropic accretion is \cite{Tarafdar}

\begin{equation}
\varepsilon = -\frac{(\gamma-1)}{(\gamma-1-c_s^2)}\sqrt{\frac{1-2/r}{[1-\frac{\lambda^2}{r^2}(1-2/r)](1-u^2)}}
\end{equation}

Taking logarithmic differentials delivers the following expression for variation of sound speed 

\begin{equation}
\delta c_s^2 = -[\gamma-1-c_s^2][p\delta r + q\delta u^2]
\end{equation}

where the quantities p and q have been defined as

\begin{eqnarray}
p &=& \frac{1}{r_c(r_c-2)} - \frac{1}{1-\frac{\lambda^2}{r_c^2}(1-2/r_c)}\bigg(\frac{\lambda^2}{r_c^3} - \frac{3\lambda^2}{r_c^4}\bigg)\\
q &=& \frac{1}{2(1 - u_c^2)}
\end{eqnarray}

\subsubsection{Constant Height}

From before, the equation for space gradient of advective velocity is \cite{Tarafdar}

\begin{equation}
\frac{du}{dr} = \frac{c_s^2\bigg[\frac{1}{r}+\frac{1}{r^2(1-2/r)}\bigg]-f_2(r,\lambda)}{(1-c_s^2)(\frac{u}{1-u^2}) - \frac{c_s^2}{u}}
\end{equation}

Going over to the $u^2-r$ plane and decomposing into a set of parametrised autonomous dynamical equations :

\begin{eqnarray}
\frac{du^2}{d\tau} &=& 2c_s^2 f_1(r) - 2f_2(r, \lambda)\\
\frac{dr}{d\tau} &=& \frac{1-c_s^2}{1-u^2} - \frac{c_s^2}{u^2}
\end{eqnarray}

where $f_1(r)$ has been defined as $f_1(r) = \frac{1}{r} + \frac{1}{r^2(1-2/r)}$

As before, the critical point conditions corresponding to standing equilibrium points of the system come out to be 

\begin{eqnarray}
[u = c_s]_{r_c} \hspace{1 cm} c_s = \sqrt{\frac{f_2}{\frac{1}{r_c} + \frac{1}{r_c^2(1-2/r_c)}}}
\end{eqnarray}

Now perturbing the system linearly about its critical points and applying the critical point conditions, gives the following coupled linear dynamical system :

\begin{eqnarray}
\frac{d(\delta u^2)}{d\tau} &=& -2f_1 q(\gamma-1-c_{sc}^2)\delta u^2 + [2c_{sc}^2 f_1'-2f_2' - 2f_1 p(\gamma -1-c_{sc}^2)]\delta r \\
\frac{d(\delta r)}{d\tau} &=& [(\gamma - 1 -c_{sc}^2)q+1]\bigg[\frac{1}{(1-u_c^2)}+\frac{1}{u_c^2}\bigg]\delta u^2 + (\gamma-1-c_{sc}^2)\bigg[\frac{1}{(1-u_c^2)}+\frac{1}{u_c^2}\bigg]p\delta r
\end{eqnarray}

For exponentially growing solution with eigenvalues $\Omega$, of the kind $\delta u^2 \sim exp (\Omega \tau)$ and $\delta r \sim exp(\Omega \tau)$, the stability matrix of the above autonomous system gives eigenvalues

\begin{equation}
\Omega^2 = \frac{2}{u_c^2(u_c^2 - 1)}[f_1 p (1 + c_{sc}^2 -\gamma) + (f_2' - c_{sc}^2f_1')\{q(1+c_{sc}^2-\gamma)-1\}]
\end{equation}

The nature of the critical point is indicated by the sign of $\Omega^2$. For $\Omega^2 < 0$, the point is of saddle type, while $\Omega^2 > 0$ indicates a centre type point. The configuration most relevant to our study is of saddle-centre-saddle type, as will be extolled upon later. This indicates a crossing of signs in the multi-critical 3-root zone of $\Omega^2$ space should be obtained, which is exactly what follows in the results.

\subsubsection{Disc in Conical Equilibrium} 

As before, \cite{Tarafdar}

\begin{equation}
\frac{du}{dr} = \frac{c_s^2\bigg[\frac{2}{r} + \frac{1}{r^2(1-2/r)}\bigg]-f_2(r,\lambda)}{\frac{u(c_s^2-1)}{(1-u^2)}-\frac{c_s^2}{u}}
\end{equation}

The parametrised autonomous system in the $u^2-r$ plane becomes 

\begin{eqnarray}
\frac{du^2}{d\tau} &=& 2c_s^2 f_1(r) - 2 f_2(r, \lambda) \\
\frac{dr}{d\tau} &=& \frac{1-c_s^2}{1-u^2} - \frac{c_s^2}{u^2}
\end{eqnarray}

where $f_1(r) = \frac{2}{r} + \frac{1}{r^2(1-2/r)}$\\

Evidently, the critical point conditions again come out to be 

\begin{eqnarray}
[u = c_s]_{r_c} \hspace{1 cm} c_s = \sqrt{\frac{f_2}{\frac{2}{r_c} + \frac{1}{r_c^2(1-2/r_c)}}}
\end{eqnarray}

Finally, the linearly perturbed coupled dynamical system becomes 

\begin{eqnarray}
\frac{d(\delta u^2)}{d\tau} &=& -2 f_1 q (\gamma -1-c_{sc}^2)\delta u^2 + [2 c_{sc}^2 f_1' - 2f_2' -2f_1 p(\gamma -1-c_{sc}^2)]\delta r \\
\frac{d(\delta r)}{d\tau} &=& [(\gamma -1-c_{sc}^2)q +1]\bigg[\frac{1}{u_c^2}+\frac{1}{(1-u_c^2)}\bigg]\delta u^2 + (\gamma -1-c_{sc}^2)\bigg[\frac{1}{u_c^2}+\frac{1}{(1-u_c^2)}\bigg]p\delta r
\end{eqnarray}

giving the eigenvalues 

\begin{equation}
\Omega^2 = \frac{2}{u_c^2(u_c^2-1)}[f_1 p (1+c_{sc}^2 -\gamma)+(f_2'-c_{sc}^2 f_1')\{q(1+c_{sc}^2 -\gamma)-1\}]
\end{equation}

\subsubsection{Flow in Vertical Equilibrium}

From before, the space gradient of advective velocity is \cite{Tarafdar}

\begin{equation}
\frac{du}{dr} = \frac{2\frac{c_s^2}{\gamma+1}f_1(r,\lambda)-f_2(r,\lambda)}{\frac{u}{1-u^2}-\frac{2c_s^2}{u(\gamma+1)}}
\end{equation}

where $f_1(r,\lambda) = \frac{2}{r}+\frac{\lambda^2}{r^3}\bigg[\frac{1-3/r}{1-\frac{\lambda^2}{r^2}(1-2/r)}\bigg]$\\

Defining $\beta^2 = \frac{2}{\gamma+1}$ and parametrising,

\begin{equation}
\frac{du^2}{d\tau} = 2\beta^2c_{sc}^2f_1 -2f_2
\end{equation}

\begin{equation}
\frac{dr}{d\tau}=\frac{1}{1-u_c^2} - \frac{\beta^2c_{sc}^2}{u_c^2}
\end{equation}

which again deliver the critical point conditions

\begin{equation}
u_c=\sqrt{\frac{1}{1+\frac{1}{\beta^2 c_{sc}^2}}} = \sqrt{\frac{f_1}{f_1 + f_2}}
\end{equation}

giving the linearly perturbed coupled dynamical system about the critical points

\begin{eqnarray}
\frac{d(\delta u^2)}{d\tau} &=& -2(\gamma -1-c_{sc}^2)\beta^2f_1 q \delta u^2 + [2\beta^2c_{sc}^2f_1'-2f_2'-2(\gamma-1-c_{sc}^2)\beta^2f_1 p]\delta r \\
\frac{d(\delta r)}{d\tau} &=& \bigg[\frac{1}{(1-u_c^2)^2}+\frac{\beta^2 c_{sc}^2}{u_c^4}+\frac{\beta^2 q (\gamma-1-c_{sc}^2)}{u_c^2}\bigg]\delta u^2 + \frac{p \beta^2(\gamma-1-c_{sc}^2)}{u_c^2}\delta r \\
\end{eqnarray}

Thereafter, the eigenvalues come out to be :

\begin{equation}
\Omega^2=-2f_1 pq \beta^4 \frac{(\gamma-1-c_{sc}^2)^2}{u_c^2}-\bigg[\frac{1}{(1-u_c^2)^2}+\frac{\beta^2 c_{sc}^2}{u_c^4}+\frac{\beta^2 q(\gamma-1-c_{sc}^2)}{u_c^2}\bigg][2f_1 p \beta^2 (1+c_s^2-\gamma)+2c_{sc}^2\beta^2 f_1' -2f_2']
\end{equation}

\subsection{Isothermal Accretion}

The conserved first integral of motion for isothermal accretion is \cite{Tarafdar}

\begin{equation}
\xi= \frac{r^2(r-2)}{[r^3-(r-2)\lambda^2](1-u^2)}\rho^{2c_s^2}
\end{equation}

which gives the space gradient of density

\begin{equation}
\frac{1}{\rho}\frac{d\rho}{dr} = \Bigg\{\frac{3r^2-\lambda^2}{[r^3-(r-2)\lambda^2]}-\frac{1}{(1-u^2)}\frac{du^2}{dr}-\frac{1}{(r-2)}-\frac{2}{r} \Bigg\}\frac{1}{2c_s^2}
\end{equation}

Note must be made here that now the sound speed is a global constant over space, hence plays no part in the variation.

\subsubsection{Constant Height Model}

For constant height model, the constant mass accretion rate is

\begin{equation}
\dot{M}=\frac{\rho u \sqrt{1-2/r}}{\sqrt{1-u^2}}r H
\end{equation}

Further differentiation of $\dot{M}$ gives a second relation for space gradient of density which on combining with (62) and parametrising in the independent parameter $\tau$, gives the following autonomous dynamical system

\begin{eqnarray}
\frac{du^2}{d\tau}&=& u^2(1-u^2)\bigg[\frac{3r^2-\lambda^2}{r^3-(r-2)\lambda^2} - \frac{1}{r-2}-\frac{2}{r}+2c_{s}^2 \Bigg\{\frac{1}{r}+\frac{1}{r^2(1-2/r)}\Bigg\}\bigg] \\
\frac{dr}{d\tau}&=&[c_s^2-u^2]
\end{eqnarray}

Here again, from above, at critical point,

\begin{equation}
[u=c_s]_{r_c}
\end{equation}

Now the linear perturbation scheme 

\begin{equation*}
u^2 = u_c^2 +\delta u^2
\end{equation*}

\begin{equation*}
r=r_c+\delta r
\end{equation*}

is applied about the critical points, to obtain the following coupled system

\begin{equation}
\begin{split}
\frac{d(\delta u^2)}{d\tau}&=(1-c_{sc}^2)c_{sc}^2\bigg[\Bigg\{\frac{6r_c}{[r_c^3-(r_c-2)\lambda^2]}-\frac{(3r_c^2-\lambda^2)^2}{[r_c^3-(r_c-2)\lambda^2]^2}+\frac{1}{(r_c-2)^2}+\frac{2}{r_c^2}\Bigg\} \\ &-2c_{sc}^2\Bigg\{\frac{1}{r_c^2}+\frac{1}{(r_c^2-2r_c)^2}\Bigg\}\bigg]\delta r 
\end{split}
\end{equation}

\begin{equation}
\frac{d(\delta r)}{d\tau}=-\delta u^2
\end{equation}

The eigenvalues for this system come out to be 

\begin{equation}
\begin{split}
\Omega^2&=-(1-c_{sc}^2)c_{sc}^2\bigg[\Bigg\{\frac{6r_c}{[r_c^3-(r_c-2)\lambda^2]}-\frac{(3r_c^2-\lambda^2)^2}{[r_c^3-(r_c-2)\lambda^2]^2}+\frac{1}{(r_c-2)^2}+\frac{2}{r_c^2}\Bigg\} \\
&-2c_{sc}^2\Bigg\{\frac{1}{r_c^2}+\frac{1}{(r_c^2-2r_c)^2}\Bigg\}\bigg]
\end{split}
\end{equation}

\subsubsection{Disc in Conical Equilibrium}

For mass accretion rate $\dot{M}=\Lambda \rho u\sqrt{\frac{1-2/r}{1-u^2}}r^2$ $(\Lambda=const.)$ and identical treatment of the above section,

\begin{eqnarray}
\frac{du^2}{d\tau}&=& u^2(1-u^2)\bigg[\frac{(3r^2-\lambda^2)}{[r^3-(r-2)\lambda^2]}-\frac{1}{(r-2)}-\frac{2}{r}+2c_s^2\Bigg\{\frac{2}{r}+\frac{1}{r(r-2)}\Bigg\}\bigg] \\
\frac{dr}{d\tau}&=& c_s^2-u^2
\end{eqnarray}

Again, $[u=c_s]_{r_c}$\\

And linear perturbation gives

\begin{equation}
\begin{split}
\frac{d(\delta u^2)}{d\tau} &= (1-c_{sc}^2)c_{sc}^2\bigg[\bigg(\frac{6r_c}{[r_c^3-(r_c-2)\lambda^2]}-\frac{(3r_c^2-\lambda^2)^2}{[r_c^3-(r_c-2)\lambda^2]^2}+\frac{1}{(r_c-2)^2}+\frac{2}{r_c^2} \bigg) \\
&-2c_{sc}^2\bigg(\frac{2}{r_c^2}+\frac{1}{(r_c^2-2r_c)^2}\bigg)\bigg]\delta r
\end{split}
\end{equation}

\begin{equation}
\frac{d(\delta r)}{d\tau}=-du^2
\end{equation}

delivering the eigenvalues

\begin{equation}
\begin{split}
\Omega^2 &= - (1-c_{sc}^2)c_{sc}^2\bigg[\bigg(\frac{6r_c}{[r_c^3-(r_c-2)\lambda^2]}-\frac{(3r_c^2-\lambda^2)^2}{[r_c^3-(r_c-2)\lambda^2]^2}+\frac{1}{(r_c-2)^2}+\frac{2}{r_c^2} \bigg)\\
&-2c_{sc}^2\bigg(\frac{2}{r_c^2}+\frac{1}{(r_c^2-2r_c)^2}\bigg)\bigg]
\end{split}
\end{equation}

\subsubsection{Disc in Vertical Equilibrium}

For a disc in vertical equilibrium, mass accretion rate is given by

\begin{equation}
\dot{M}=\rho r^2 u c_s\sqrt{2[r^3-(r-2)\lambda^2]}
\end{equation}

which along with $\xi$, gives the following autonomous system of equations in the $u^2-r$ plane

\begin{eqnarray}
\frac{du^2}{d\tau}&=&\bigg[-\frac{(3r^2-\lambda^2)(c_s^2+1)}{[r^3-(r-2)\lambda^2]}+\frac{1}{(r-2)}+\frac{2}{r}-\frac{4c_s^2}{r} \bigg]u^2(1-u^2)\\
\frac{dr}{d\tau} &=& [-u^2+c_s^2(1-u^2)]
\end{eqnarray}

The critical point condition emerging from above is 

\begin{equation}
\bigg[u^2=\frac{c_s^2}{1+c_s^2}\bigg]_{r_c}
\end{equation}

Linear perturbation of the system gives 

\begin{equation}
\frac{d(\delta u^2)}{d\tau} = u_c^2(1-u_c^2)\bigg[-\frac{6r_c(1+c_{sc}^2)}{[r_c^3-(r_c-2)\lambda^2]}+\frac{(3r_c^2-\lambda^2)(1+c_{sc}^2)}{[r_c^3-(r_c-2)\lambda^2]^2}-\frac{1}{(r_c-2)^2}-\frac{2}{r_c^2}+\frac{4c_{sc}^2}{r_c^2} \bigg]\delta r
\end{equation}

\begin{equation}
\frac{d(\delta r)}{d\tau}=-(1+c_{sc}^2)\delta u^2
\end{equation}

which delivers the eigenvalues

\begin{equation}
\Omega^2 = u_c^2(1-u_c^2)(1+c_{sc}^2)\bigg[-\frac{6r_c(1+c_{sc}^2)}{[r_c^3-(r_c-2)\lambda^2]}+\frac{(3r_c^2-\lambda^2)(1+c_{sc}^2)}{[r_c^3-(r_c-2)\lambda^2]^2}-\frac{1}{(r_c-2)^2}-\frac{2}{r_c^2}+\frac{4c_{sc}^2}{r_c^2} \bigg]
\end{equation}

\section{Results And Discussion}
\subsection{A: The Parameter Space Dependence Of Stationary State Solutions}

As has been mentioned, the critical flow points are obtained as the stationary solutions to the flow equations of Section 2, in the velocity-radial distance phase space. They are therefore, in effect, completely tuned by the flow parameters $[\frac{\varepsilon}{T} ,\lambda ,\gamma ]$. Figures [1a] to [2c] show this dependence of the stationary state flow solutions on the flow parameters, in the parameter space spanning the entire zone of astrophysically relevant initial conditions. Figures (1a),(2a) correspond to constant height model,(1b),(2b) correspond to conical equilibrium model and (1c),(2c) correspond to vertical equilibrium model. $\gamma$ has been kept fixed at 4/3. The dependence has been shown for all three flow geometries. For the isothermal graphs, temperature T has been scaled in units of $10^{10}$K. This parameter space  classification naturally emerges as a by-result of the location of critical flow points which was  necessary for this study, and has been shown here for reasons of completeness. It has also been explored in previous works, for the general relativistic case \cite{Tarafdar,Goswami}.

\begin{figure}[H]
\begin{center}
\subfloat[Constant Height Model]{\includegraphics[width = 2.8in]{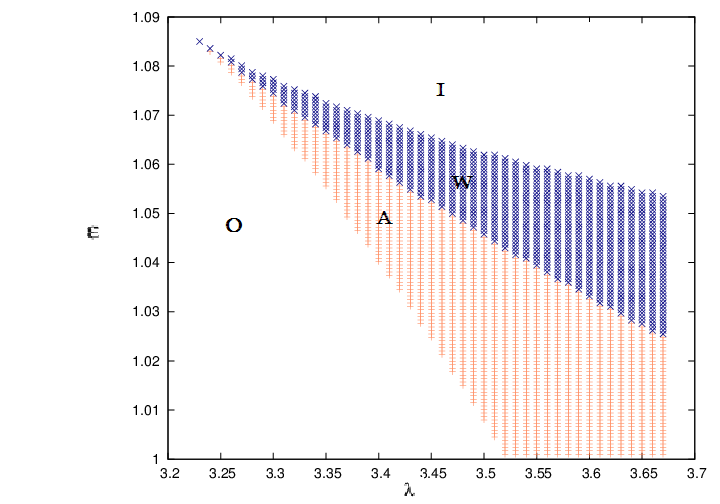}} 
\subfloat[Disc in Conical Equilibrium]{\includegraphics[width = 2.8in]{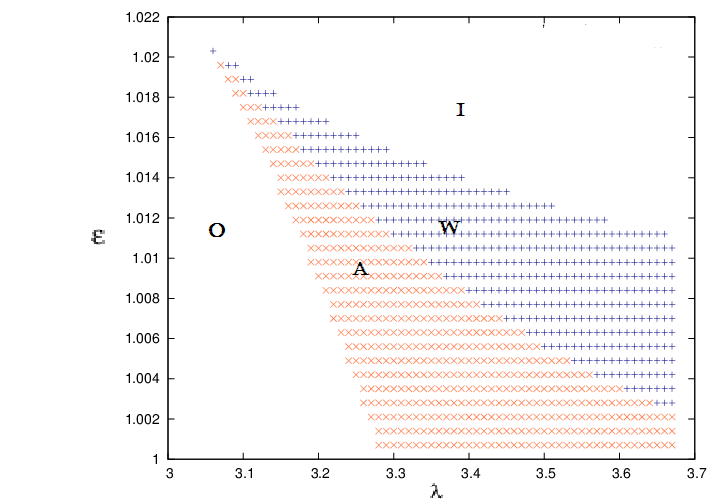}}\\
\subfloat[Disc in Vertical Equilibrium]{\includegraphics[width = 2.8in]{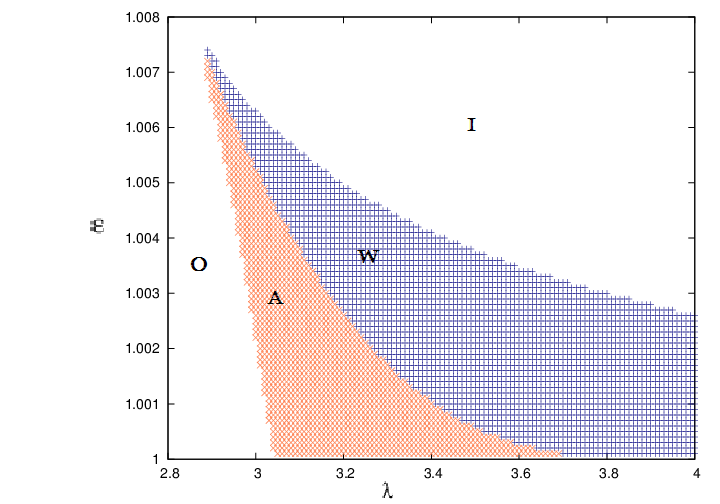}}
\end{center}
\caption{Parameter space spanned by stationary state solutions for Polytropic Case. $\gamma$ has been kept fixed at $\frac{4}{3}$. O and I denote zones where single critical points arise, at different length scales. The central coloured wedge-shaped zone spans the region where 3 critical points emerge.  A represents the Accretion Zone, W the wind zone. }
\label{some example}
\end{figure}

The regions O and I flanking the wedge shaped region correspond to zones where single critical points arise. These two zones occur in the two extreme regimes of low $[\frac{\varepsilon}{T},\lambda]$, and high $[\frac{\varepsilon}{T},\lambda]$. The physics behind them correspond to two energetically different scenarios.

\begin{figure}[H]
\begin{center}
\subfloat[Constant Height Model]{\includegraphics[width = 2.8in]{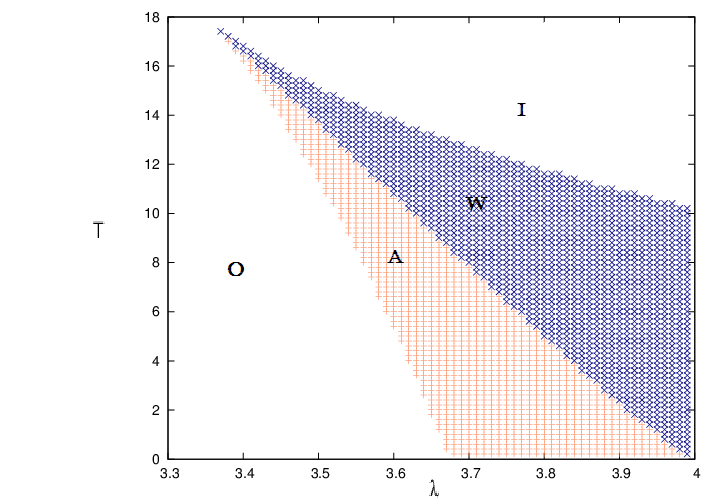}} 
\subfloat[Disc in Conical Equilibrium]{\includegraphics[width = 2.8in]{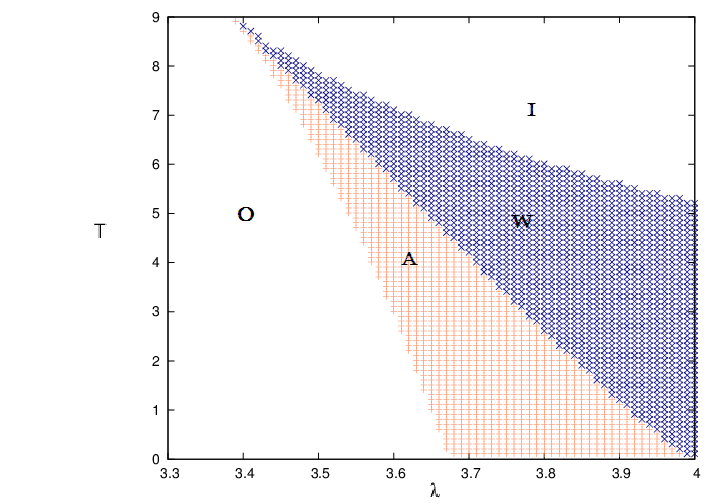}}\\
\subfloat[Disc in Vertical Equilibrium]{\includegraphics[width = 2.8in]{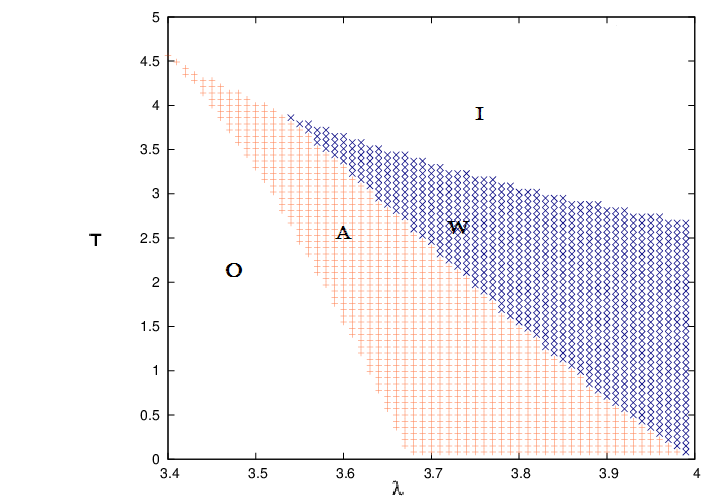}}
\end{center}
\caption{Parameter Space spanned by stationary state solutions for Isothermal Case. T has been scaled in units of $10^{10} K$. O and I denote zones where single critical points arise, at different length scales. The central coloured wedge-shaped zone spans the region where 3 critical points emerge.  A represents the Accretion Zone, W the wind zone. }
\label{some example}
\end{figure}

Region O (standing for ’Outer’) represents the region where a lone critical point is formed, at very large distances from the event horizon. This is the case of low $\frac{\varepsilon}{T}$ and low angular momentum, or ’cold flow’. The physical picture is easy to grasp. The flow starts out from the matter source very large length scales, towards the accretor , driven by its gravitational
field. Unlike the case of spherical Bondi flow\cite{Bon52}, the growth of the advective velocity field is not monotonically growing for axisymmetric rotating accretion flow. It is modulated by the angular momentum, which sets up a centrifugal barrier that must be overcome at the cost of gravity. For this ’O’ region, the $\lambda$ and $\frac{\varepsilon}{T}$ values are both low. As such, the flow is energetically too weak (or 'cold')\cite{Goswami} to counteract the effect of the growing gravity field for very long, which is hence able to overcome them and allow the flow to achieve criticality even at points quite early on in its trajectory, when the fluid is still far from the event horizon.\\

A complete reversal of the above conditions corresponds to the region I, the erstwhile ’Inner’ region or the region of ’hot’ flow. Here a single critical point is formed very close to the event horizon. $\lambda$ as well as $\frac{\varepsilon}{T}$ are now quite high. The accreting matter now needs to travel a large distance before the gravity field garners sufficient strength to overcome the resistance put up by $\lambda$ and the advective velocity can catch up with the sonic velocity to achieve criticality.\\

The most striking non-trivial feature of this flow is the region of multicriticality (a three critical point zone), demarcated by the wedge shaped zone in the figures. As it has emerged from the results of the next section, the only configuration of these three critical points that is obtained in this analysis is of the saddle-centre-saddle type, which is to be expected for real, physical transonic solutions connecting infinity with the event horizon. This region has been further divided into regions marked A (accretion) and W (wind). The mathematical basis for this separation, as has already been stated earlier, is the entropy accretion rate $\dot{\Xi}$. The region A corresponds to regions where $\dot{\Xi}_{in}$ > $\dot{\Xi}_{out}$. The W zone is for regions with $\dot{\Xi}_{in}$ < $\dot{\Xi}_{out}$. The boundary zone between the accretion and wind zones is where the degeneracy $\dot{\Xi}_{in}$ = $\dot{\Xi}_{out}$ holds. This degenerate locus corresponds to a heteroclinic trajectory \cite{Jordan}, which are notoriously unstable, and will blossom out into various kinds of bifurcations on the slightest tweaking of parameters \cite{Goswami}. It is also reiterated here that multicriticality must not automatically be taken to guarantee multitransonicity, unless some mechanism like shock transitions are possible and accessible to the system. A true multi-transonic solution can only be realised for shock solutions, if the criteria for energy preserving relativistic Rankine-Hugoniot shock for polytropic accretion\cite{Ran70,Hugo871,Hugo873,Landau} and temperature-preserving relativistic shock for isothermal accretion are met\cite{Tarafdar,Yang95,Yuan96}.

\subsection{B: A Dynamical Systems Study Of The Fixed Points}

The eigenvalues $\Omega^2$ of the stability matrix corresponding to the coupled, linearized, autonomous dynamical system of the flow equations in the $u^2-r$ plane have been derived earlier in Section[3], for all the three flow geometries. This has given $\Omega^2$ as functions of the flow parameters ${\frac{\varepsilon}{T},\lambda,\gamma}$ and the critical point coordinates $r_c$. Keeping $\gamma$ fixed at 4/3, and using the values of critical points obtained from computation of Section [2], the surface plots giving the dependence of $\Omega^2$ on $\frac{\varepsilon}{T}$, and $\lambda$ have been shown in figures [3a] to [8c]. The label [a] indicates a plot for a inner critical point, [b] for middle critical point, and [c] for outer critical point. Two different colours have been used, to indicate the accretion and wind zones (navy for wind, rust for accretion). In this section too, temperature has been scaled in units of $10^{10}$.\\

From a knowledge of dynamical systems, the sign of the value of $\Omega^2$ indicates the nature of the critical point. Positive values indicate a saddle type point, while negative values indicate a centre type point. It is clearly evident from the plots that the inner and outer critical points for all the flow geometries (under both polytropic and isothermal accretion) are saddle type points, flanking a centre type middle critical point. This is the only possible configuration that can allow real, physical transonic accretion flow trajectories to connect the source at infinity with the accretor via a path that can double up on itself, i.e a homoclinic path.\\

A few more remarks are in order on the surface plots. The absolute value of $\Omega^2$ is seen to anti-correlate with the critical point coordinates, as is evident from its values coming out to be largest in the plots for the inner critical point, verifying this claim.  $\Omega^2$ values for the inner and outer saddle type points differ by several orders of magnitude. Examining the dependence on flow parameters for these two saddle type points (inner and outer), the dependence for the inner point is seen to be much weaker than the outer one. Variation of $\Omega^2$ for the inner critical point shows very slight inverse growth dependence on $\varepsilon$, and slow positive growth dependence on $\lambda$. Whereas for the outer critical point, variation of $\Omega^2$ with angular momentum shows sudden, rapid rise. The change is not so marked, however, for variation with respect to flow energy $\varepsilon$ over its entire range of variation. Another point to note is that the $\Omega^2$ surfaces show opposite curvatures for these two cases.\\

\begin{figure}[H]
\begin{center}
\subfloat[$\Omega^2$ for inner critical point. ]{\includegraphics[width = 3.5in]{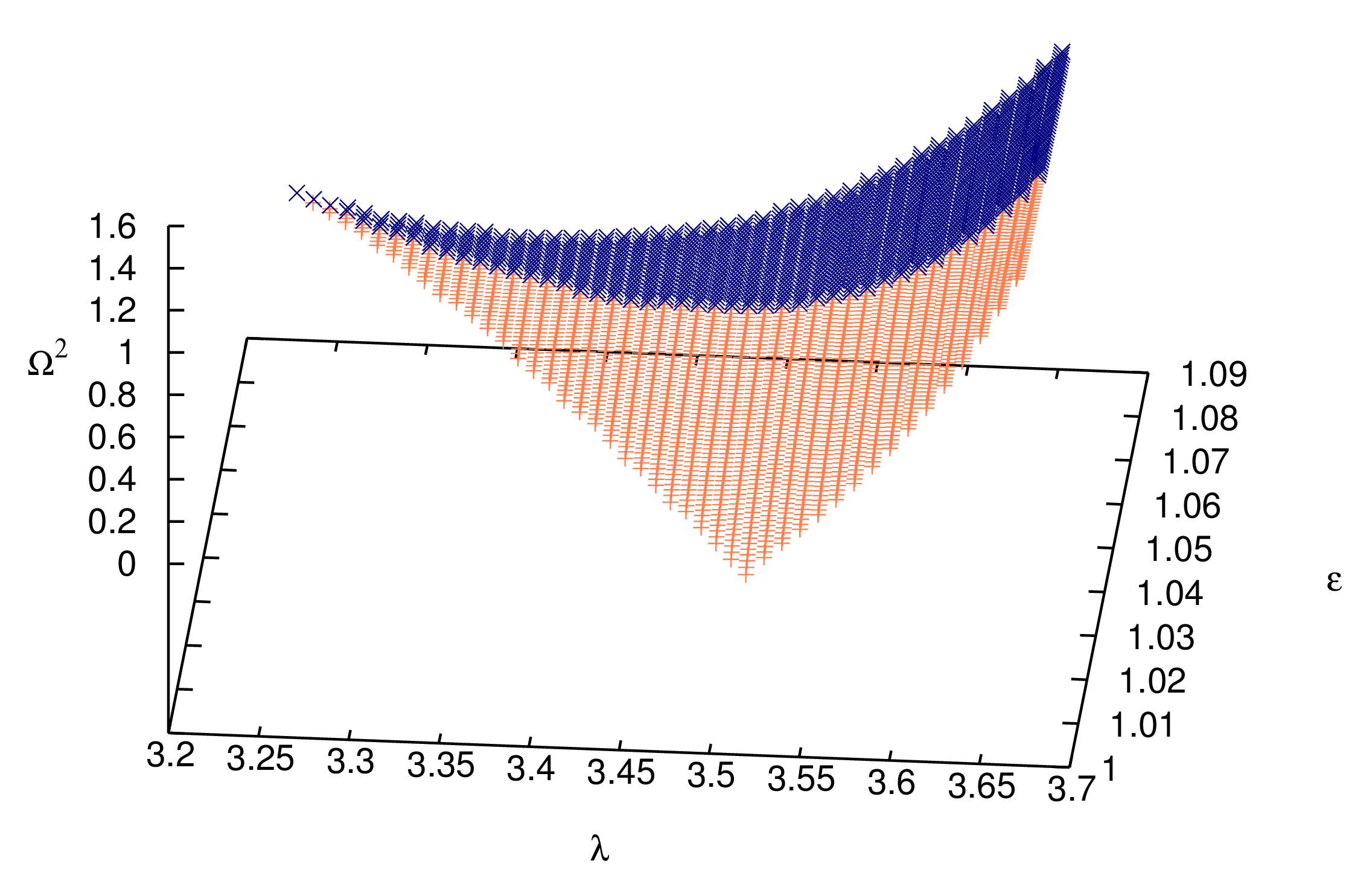}}\\ 
\subfloat[$\Omega^2$ for middle critical point]{\includegraphics[width = 3.5in]{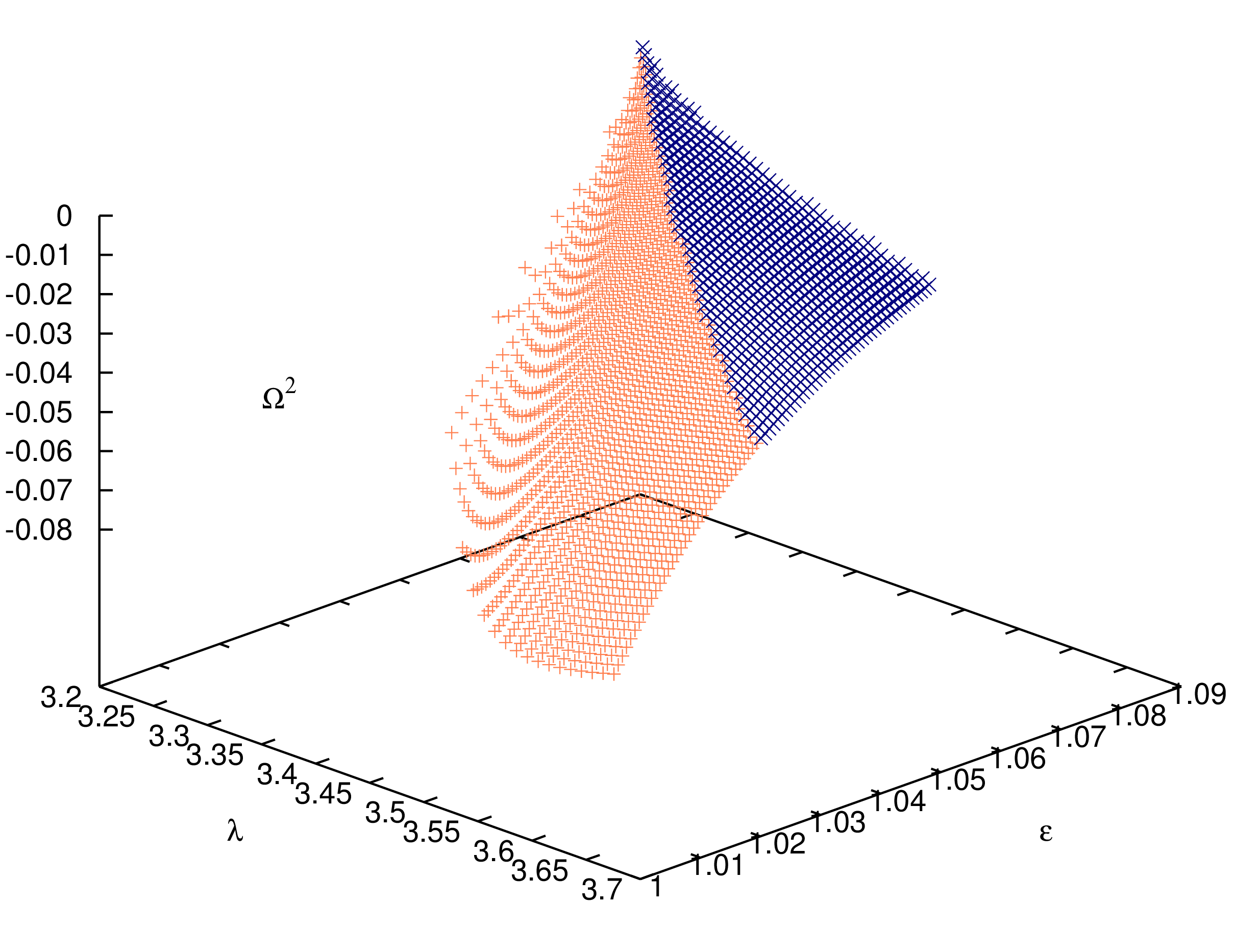}}\\
\subfloat[$\Omega^2$ for outer critical point]{\includegraphics[width = 3.5in]{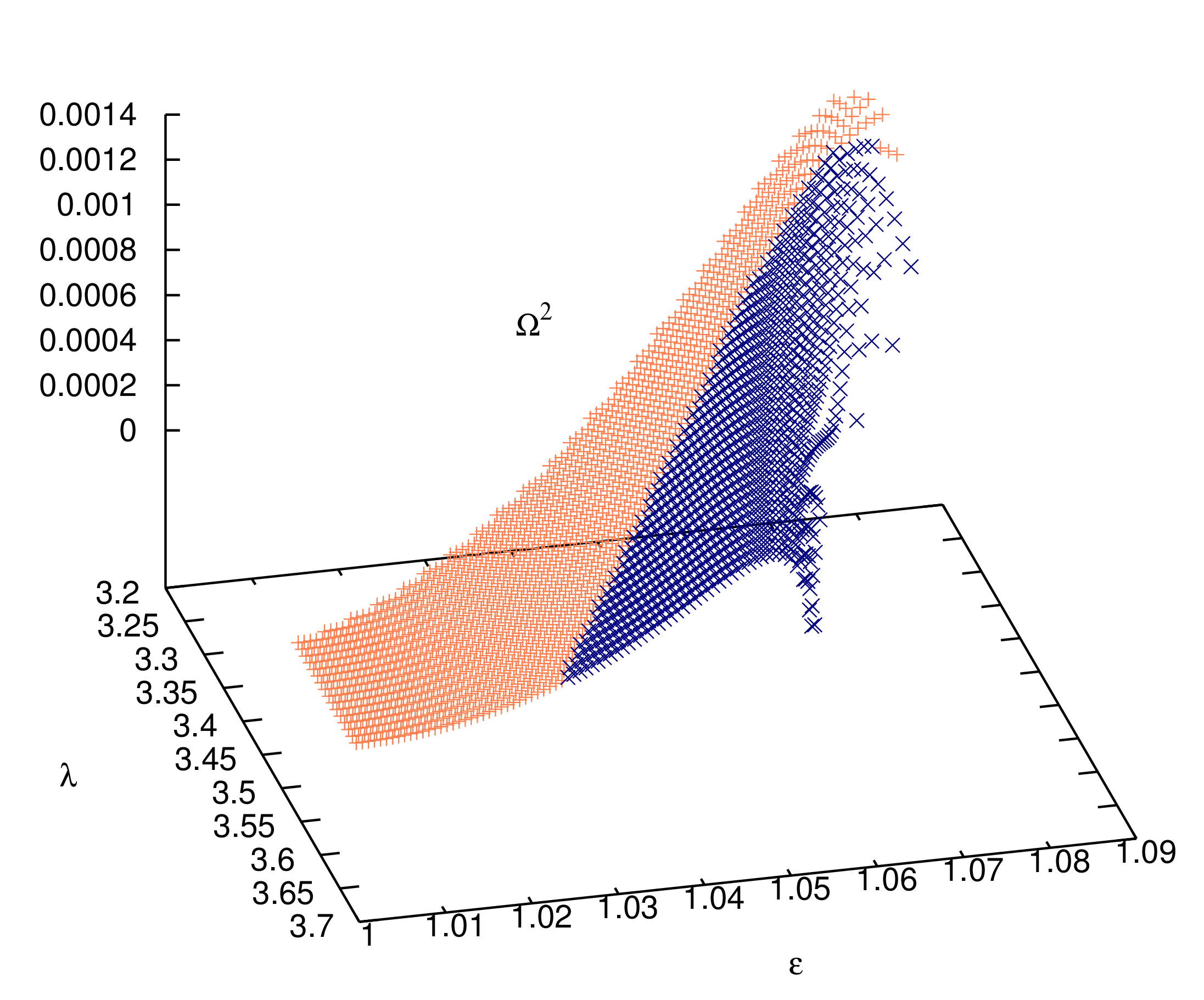}}
\end{center}
\caption{$\Omega^2$ surface plots for Polytropic Case - Constant Height Model. \\
\hspace{1.4 cm}Colour scheme - rust denotes Accretion, navy denotes Wind.}
\label{some example}
\end{figure}

\begin{figure}[H]
\begin{center}
\subfloat[$\Omega^2$ for inner critical point.]{\includegraphics[width = 3.15in]{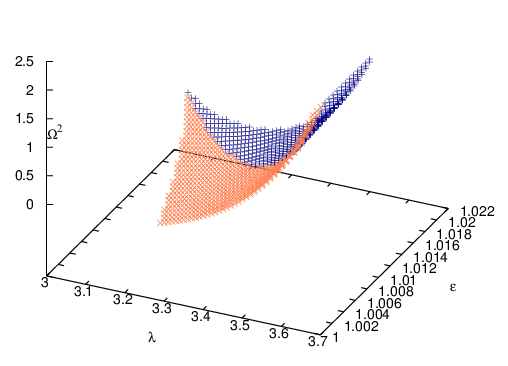}} \\
\subfloat[$\Omega^2$ for middle critical point.]{\includegraphics[width = 3.25in]{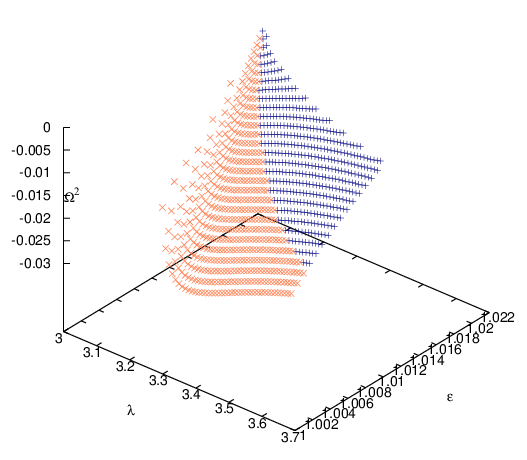}}\\
\subfloat[$\Omega^2$ for outer critical point.]{\includegraphics[width = 3.15in]{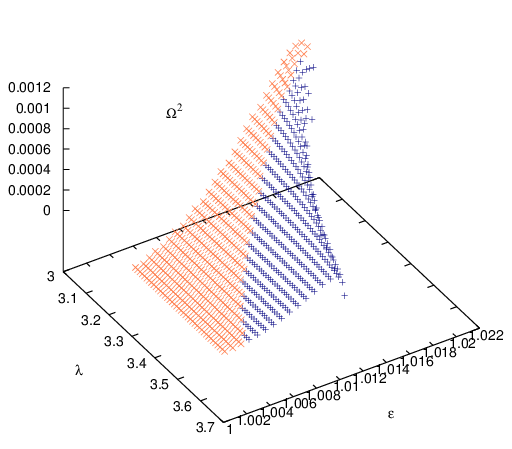}}
\end{center}
\caption{$\Omega^2$ surface plots for Polytropic Case - Conical Equilibrium Model}
\label{some example}
\end{figure}

\begin{figure}[H]
\begin{center}
\subfloat[$\Omega^2$ for inner critical point.]{\includegraphics[width = 3.5in]{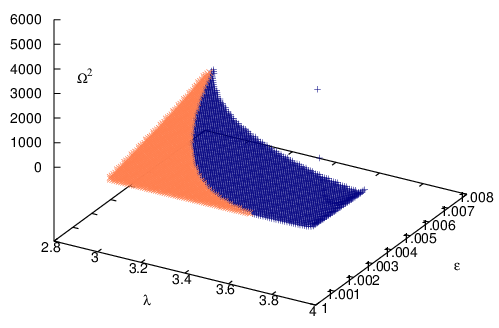}}\\
\subfloat[$\Omega^2$ for middle critical point.]{\includegraphics[width = 3.5in]{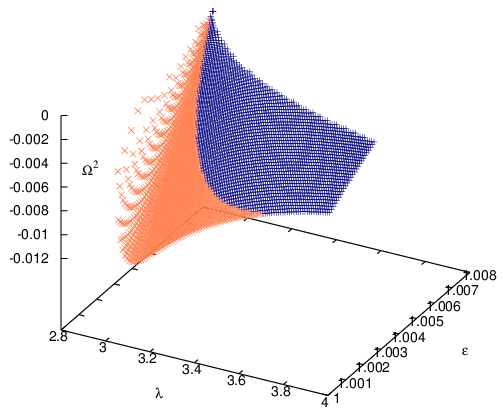}}\\
\subfloat[$\Omega^2$ for outer critical point.]{\includegraphics[width = 3.3in]{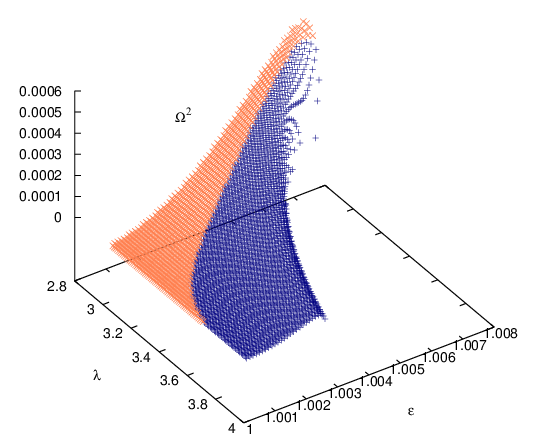}}
\end{center}
\caption{$\Omega^2$ surface plot for Polytropic Case - Vertical Equilibrium Model} 
\label{some example}
\end{figure}

\begin{figure}[H]
\begin{center}
\subfloat[$\Omega^2$ for inner critical point.]{\includegraphics[width = 3in]{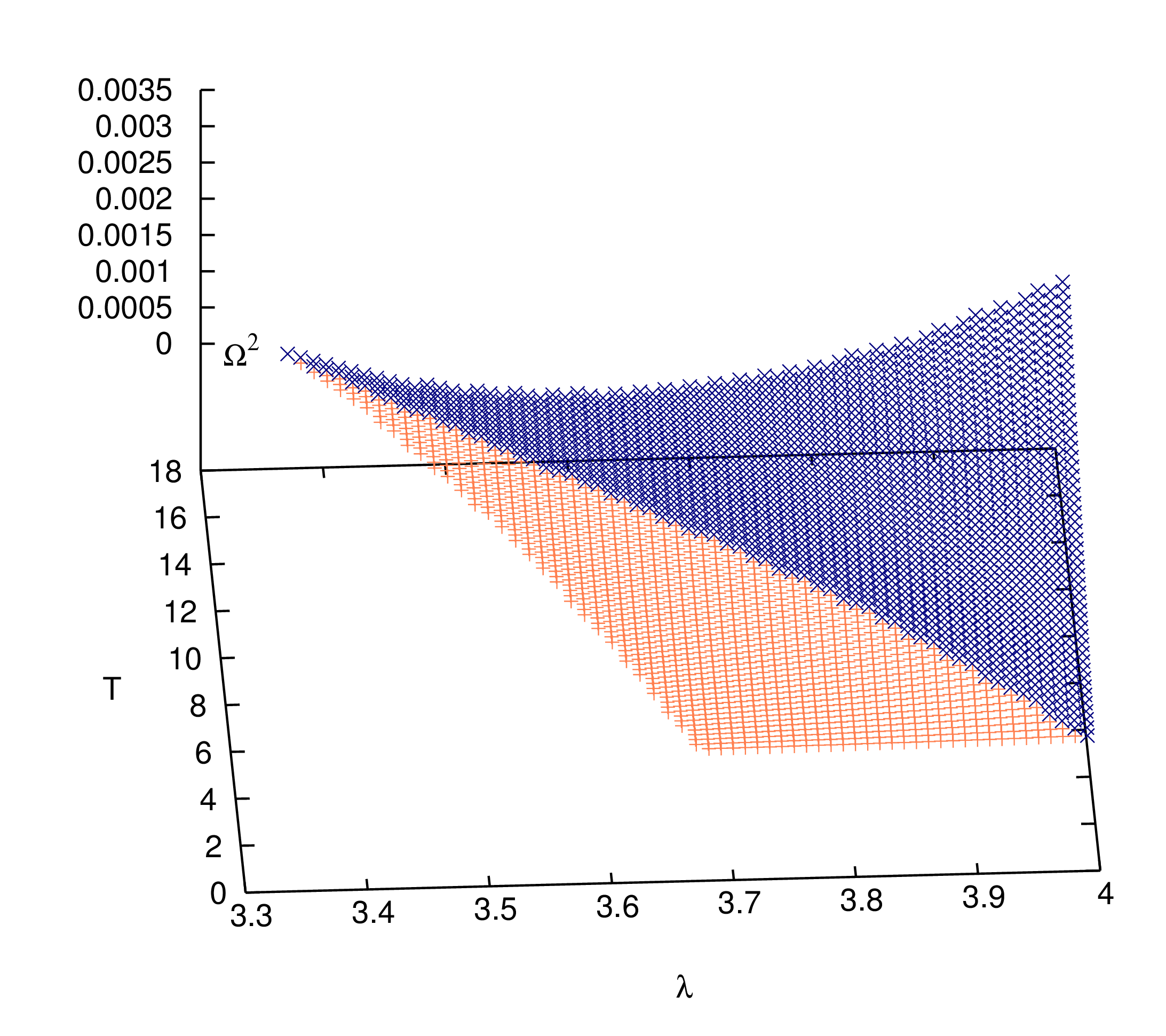}}\\ 
\subfloat[$\Omega^2$ for middle critical point.]{\includegraphics[width = 3.2in]{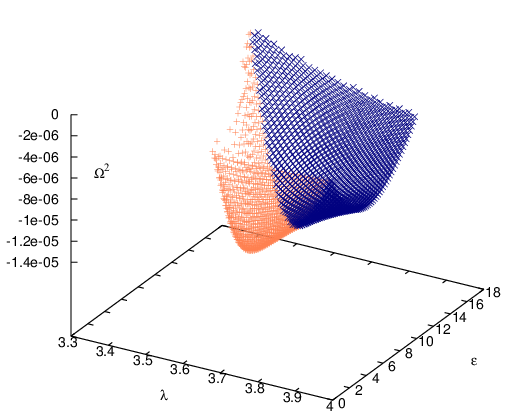}}\\
\subfloat[$\Omega^2$ for outer critical point.]{\includegraphics[width = 3.3in]{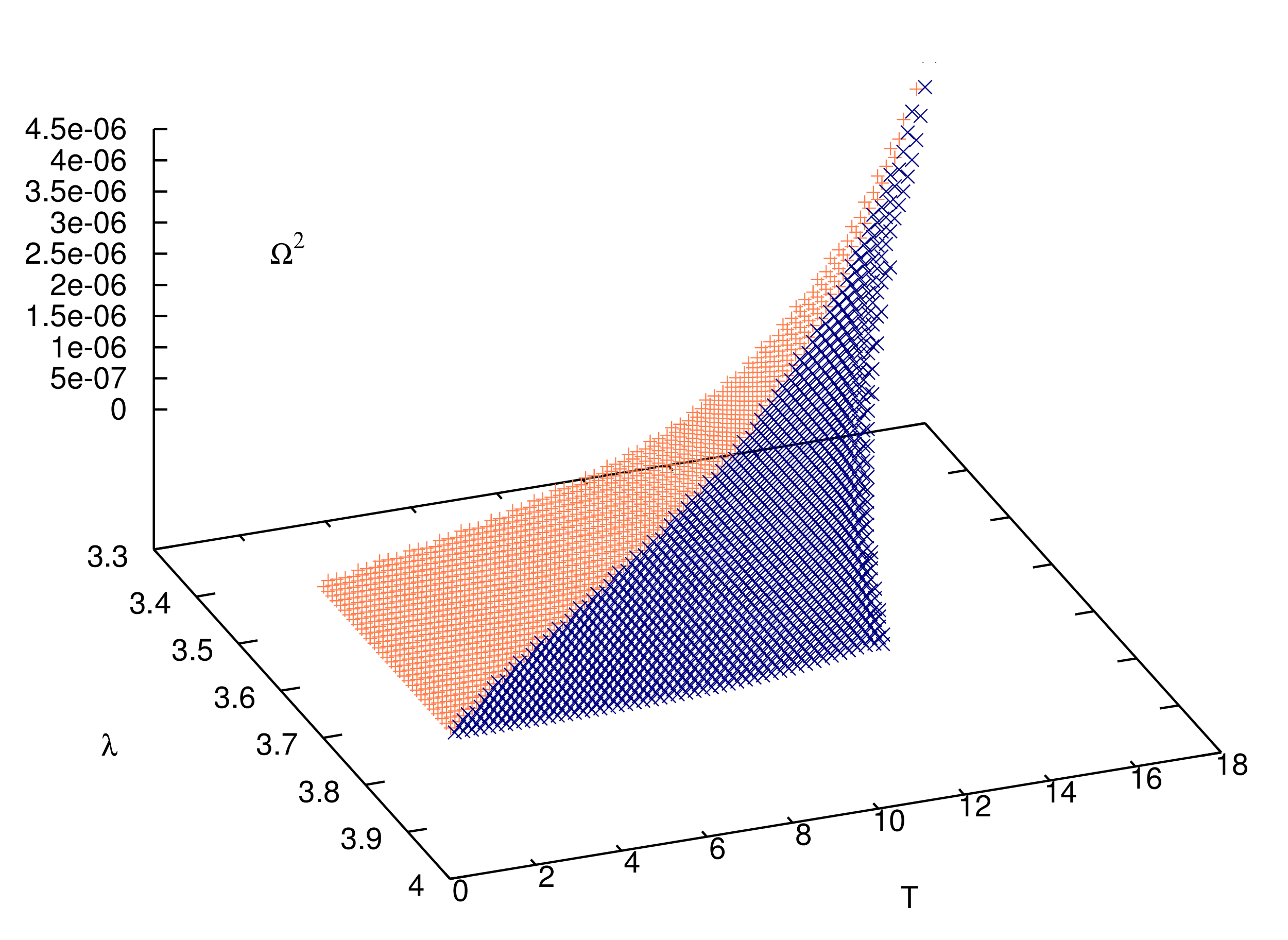}}
\end{center}
\caption{$\Omega^2$ surface plot for Isothermal Case - Constant Height Model.\\
\hspace{1.4 cm}Temperature has been scaled in units of $10^{10}K$}
\label{some example}
\end{figure}

\begin{figure}[H]
\begin{center}
\subfloat[$\Omega^2$ for inner critical point.]{\includegraphics[width = 3.2in]{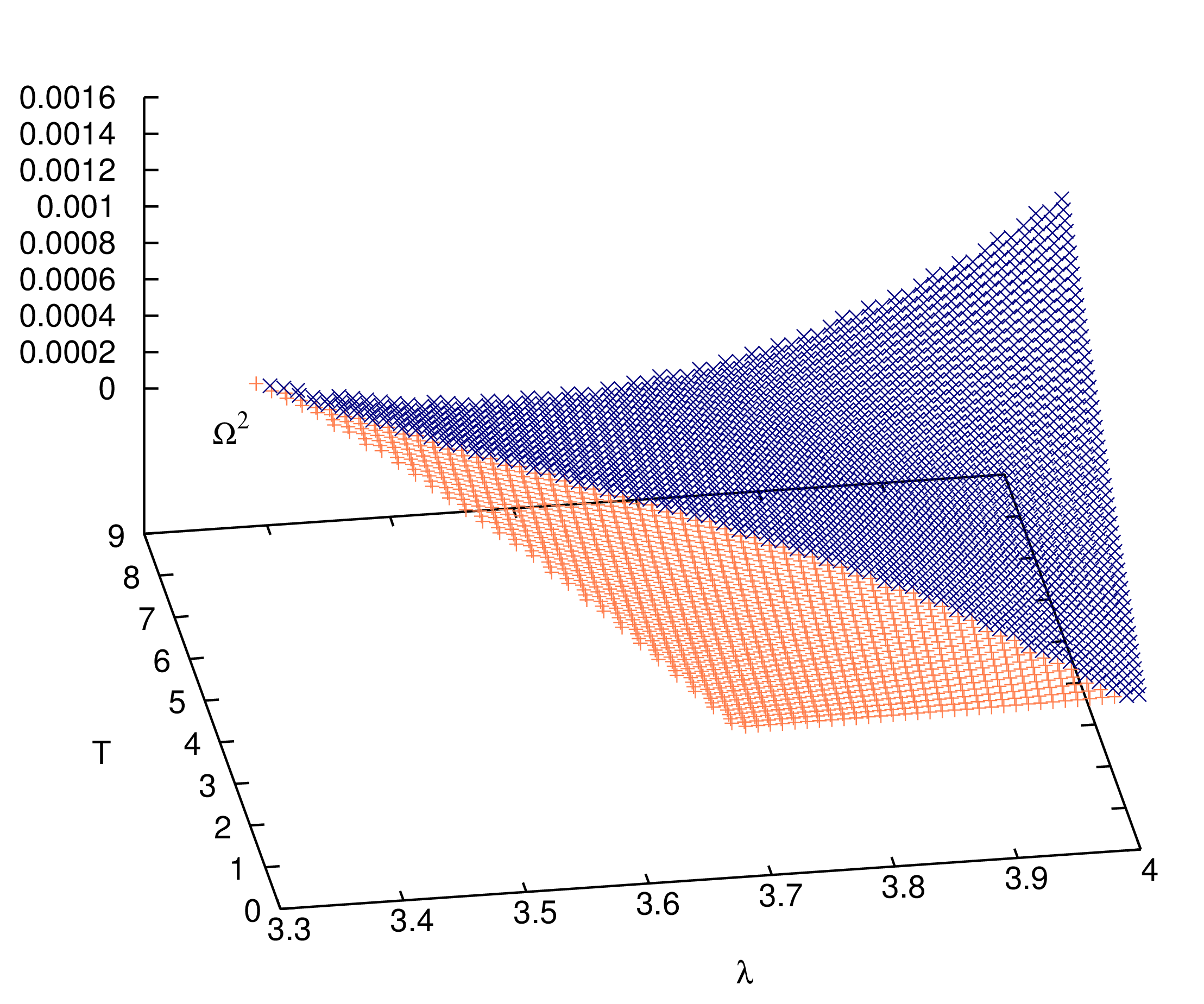}} \\
\subfloat[$\Omega^2$ for middle critical point.]{\includegraphics[width = 3.2in]{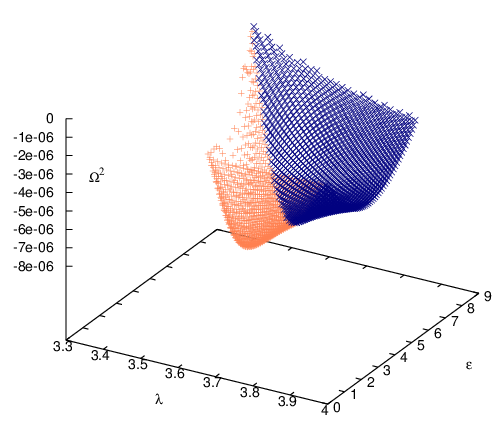}}\\
\subfloat[$\Omega^2$ for outer critical point.]{\includegraphics[width = 3.2in]{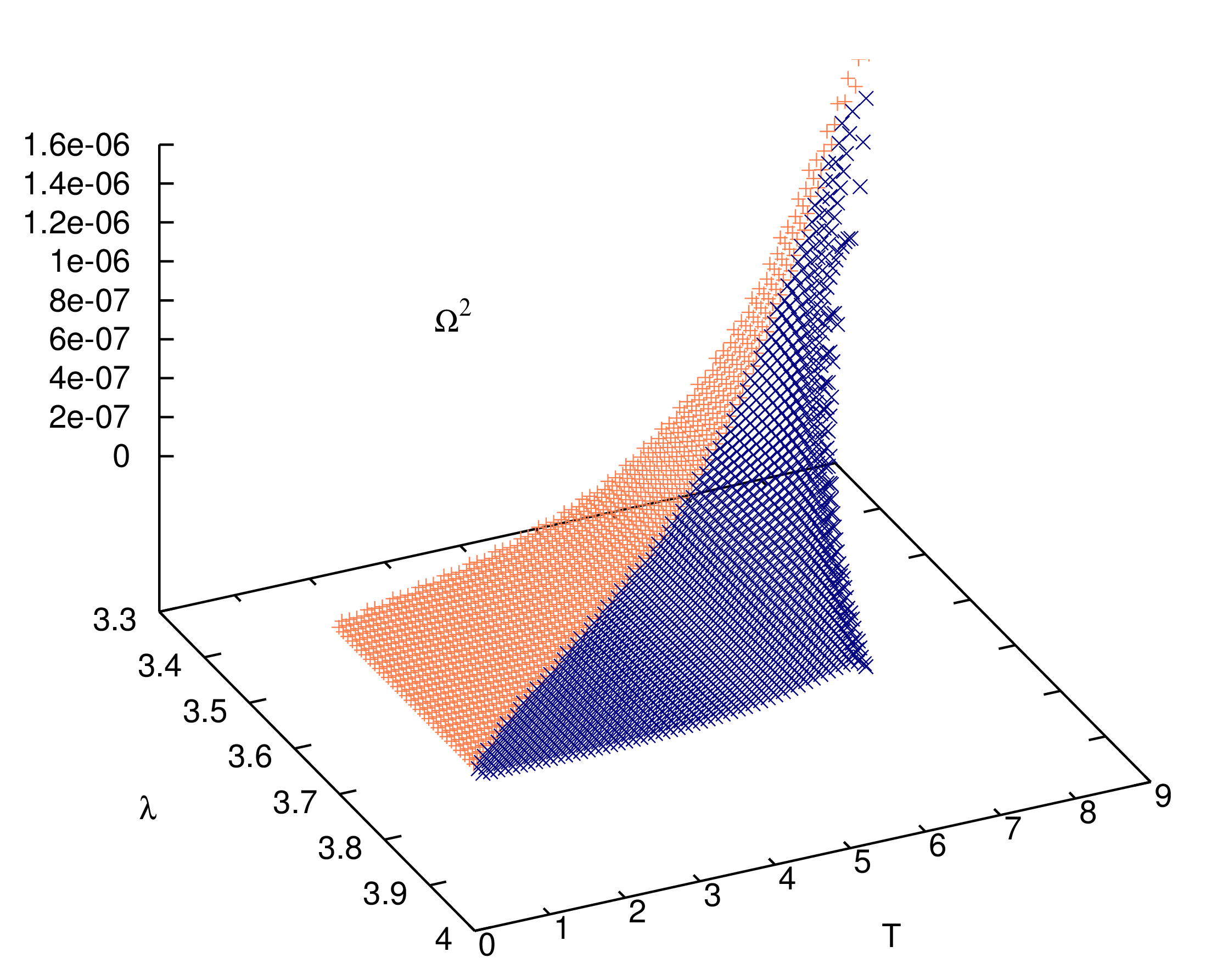}}
\end{center}
\caption{$\Omega^2$ surface plot for Isothermal Case - Conical Equilibrium Model}
\label{some example}
\end{figure}

\begin{figure}[H]
\begin{center}
\subfloat[$\Omega^2$ for inner critical point.]{\includegraphics[width = 3.5in]{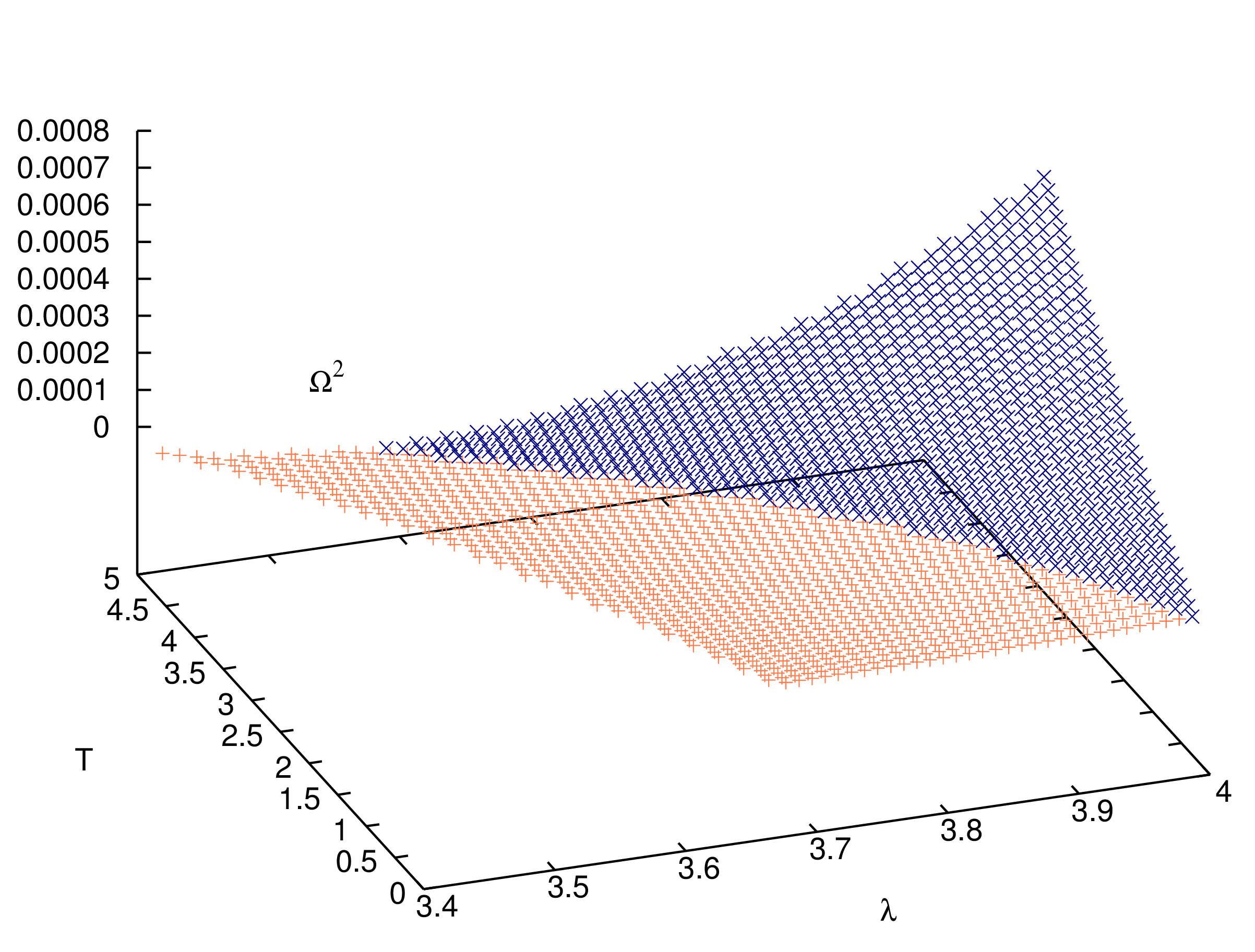}}\\ 
\subfloat[$\Omega^2$ for middle critical point.]{\includegraphics[width = 3.5in]{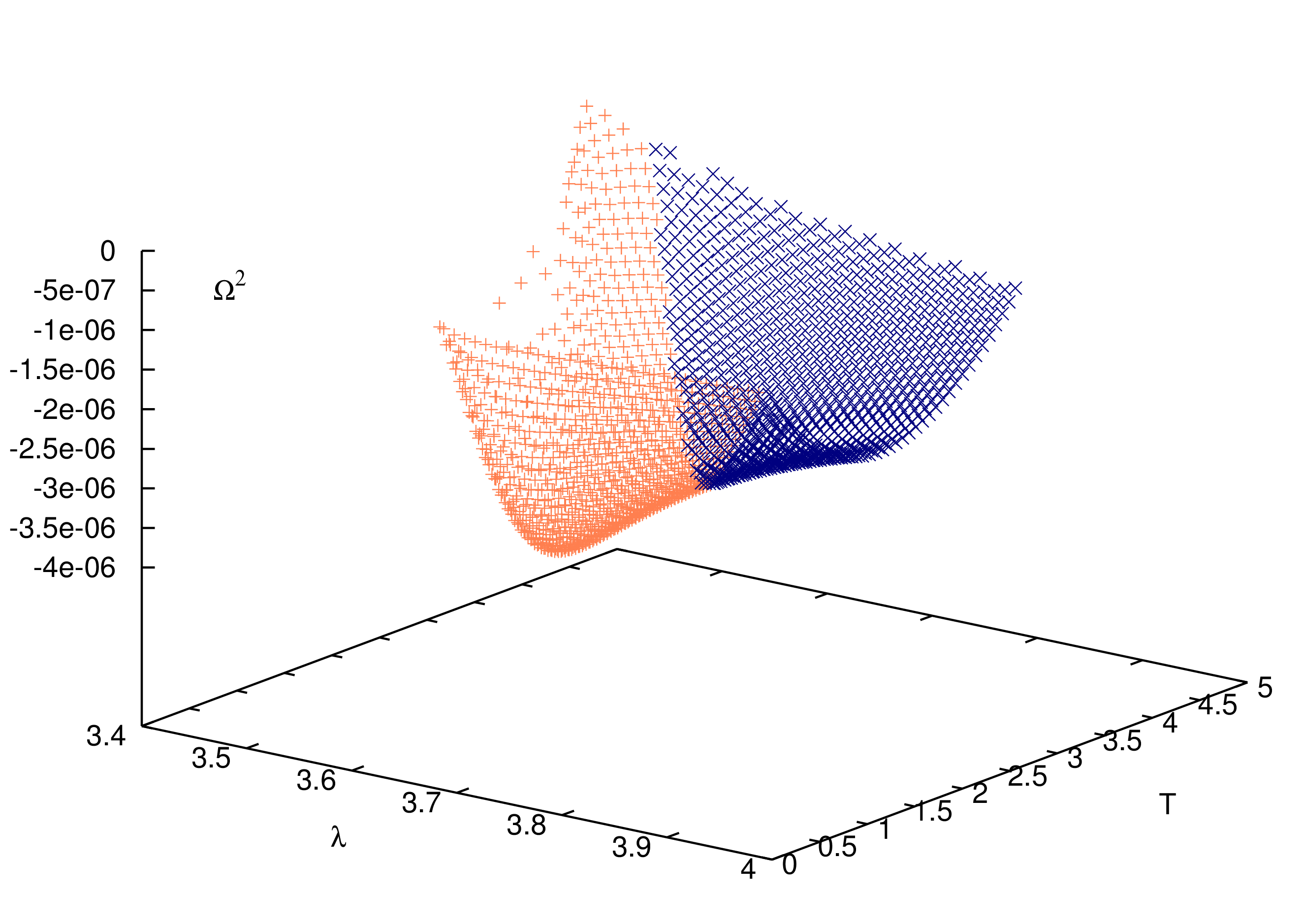}}\\
\subfloat[$\Omega^2$ for outer critical point.]{\includegraphics[width = 3.5in]{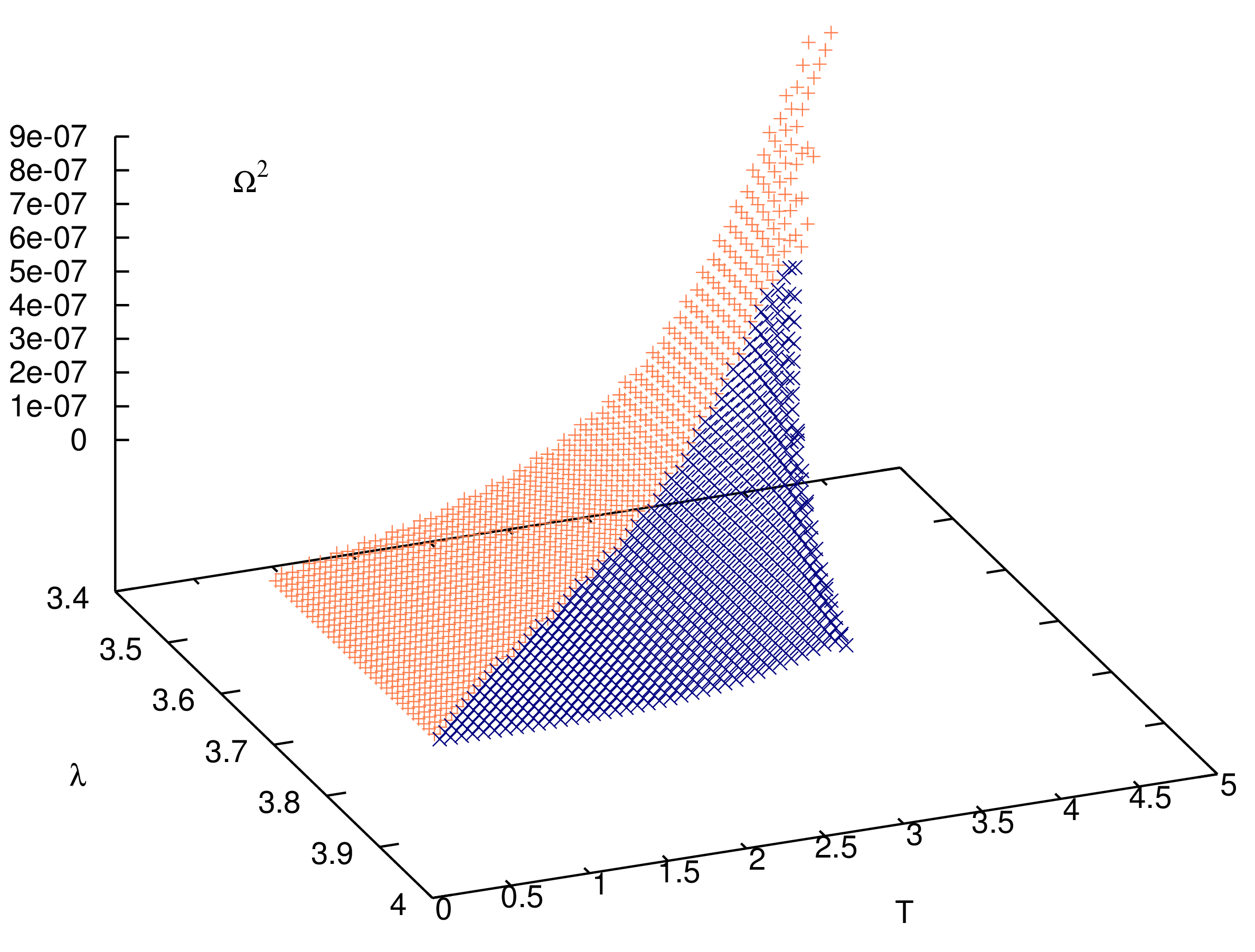}}
\end{center}
\caption{$\Omega^2$ surface plot for Isothermal Case - Vertical Equilibrium Model}
\label{some example}
\end{figure}

Figures [9a] to [14b] show the projection of $\Omega^2$ on the $\lambda$ axis, reducing another degree of freedom by now keeping $\frac{\varepsilon}{T}$ also fixed along with $\gamma$. These figures make the bifurcations happening in this entire scheme more apparent. The initial region (not shown) will be that of a single critical point of saddle type, and hence $\Omega^2$ will start out positive. At a certain value of $\lambda$, a saddle-centre type bifurcation just appears. This would correspond to a point on the junction of the O and A regions, in the parameter space. Two new critical points, a centre type and a saddle type, are born here. In $\Omega^2$ space, they correspond to branching out of negative and positive values. As $\lambda$ increases further, at a certain higher value of $\lambda$, the centre type point now coincides with one of the saddle type critical points. This merging of the centre and one of the saddle type points (the outer saddle point) corresponds to a second bifurcation in phase space. The remaining saddle point survives into the lone critical point zone. The three different branches of $\Omega^2$ corresponding to the three different critical point coordinates,have been shown in three different colours. The branches for the middle and outer critical points (shown in green and blue respectively) are very close, but it must be noted that one lies entirely in the positive quadrant, while the other lies below. It has not been possible to resolve them further while simultaneously showing all three branches on the same figure, due to difference in orders of magnitude between $\Omega^2$ values of the inner and outer saddle type critical points. The graphs for the accretion and wind zones have been shown on separate figures for each disc geometry, to skirt problems of overlap.\\

\section{Concluding Remarks}

The main focus of this study was axially symmetric multi transonic accretion in strong gravity.The time independent, general relativistic fluid equations governing the accretion flow onto a Schwarzschild black hole were studied, for three different geometric configurations of the accretion disc- the constant height, the conical equilibrium and the hydrostatically balanced vertical equilibrium models. Thermodynamic conditions ranging from variable polytropic to pure isothermal were used to describe the flow. The study was done by identifying the coupled differential equations in the flow velocity and critical radial coordinate with a first order autonomous dynamical system. This treatment necessitated locating the critical points, and identifying the parameter space spanning the initial boundary conditions. Out of the six configurations under study, the critical and sonic points were seen to coincide for four cases- in effect,  making the location of the acoustic horizon trivial. However, for the disc in vertical equilibrium, for both isothermal and polytropic flow, the critical and sonic points were seen to be non degenerate. Location of the acoustic horizon for these two cases would thus need taking recourse to explicit integration of the flow equations. Thereafter a linear variational analysis of this system was carried out around the critical points. The semi-analytical formalism developed here provided comprehensive qualitative insight into the nature of the phase trajectories of the flow, without taking recourse to explicit numerical integration. The most interesting feature that emerged from this analysis was the multi-critical zone in phase space, and its subsequent resolution into a saddle-centre-saddle type configuration or a homoclinic orbit for real, multi-transonic accretion. On another note, axisymmetric black hole accretion is repeatedly cited as an example of multi-transonic accretion. It is mentioned here again that multi-criticality must not be taken to be topologically isomorphic with multi-transonicity \cite{Tarafdar}. The possibility of shock solutions to arise in the system, must exist, where the Rankine-Hugoniot shock conditions can be satisfied at the shock front. A full treatment in the general relativistic framework, for locating these shocks, and investigating the dependence of the shock location, as well as pre- to post shock ratios of various accretion variables, on different dynamical and thermodynamic properties of the flow, remains to be seen. \\

\begin{figure}[H]
\begin{center}
\subfloat[Figure for Accretion Zone]{\includegraphics[width = 4in]{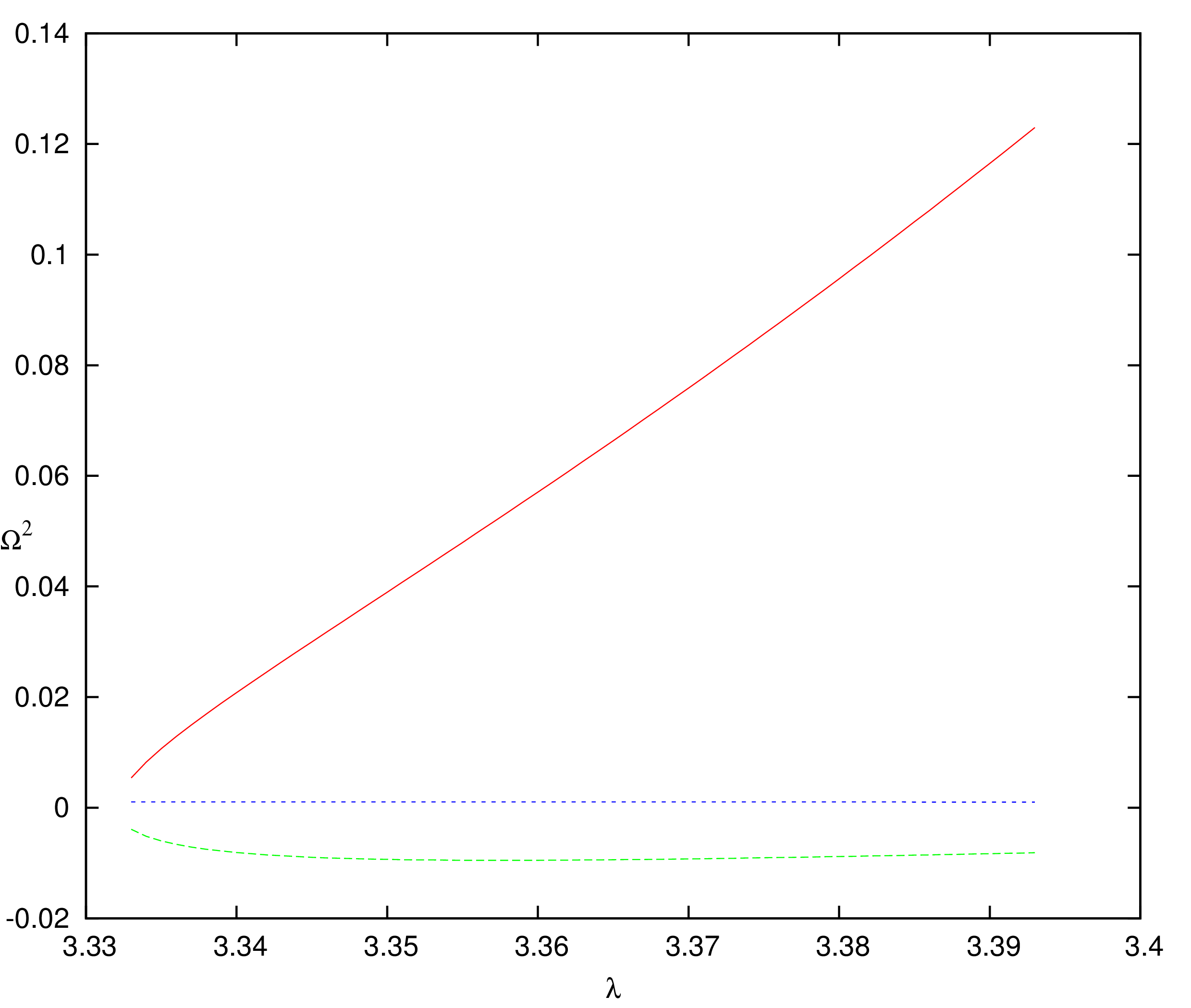}} \\
\subfloat[Figure for Wind Zone]{\includegraphics[width = 4in]{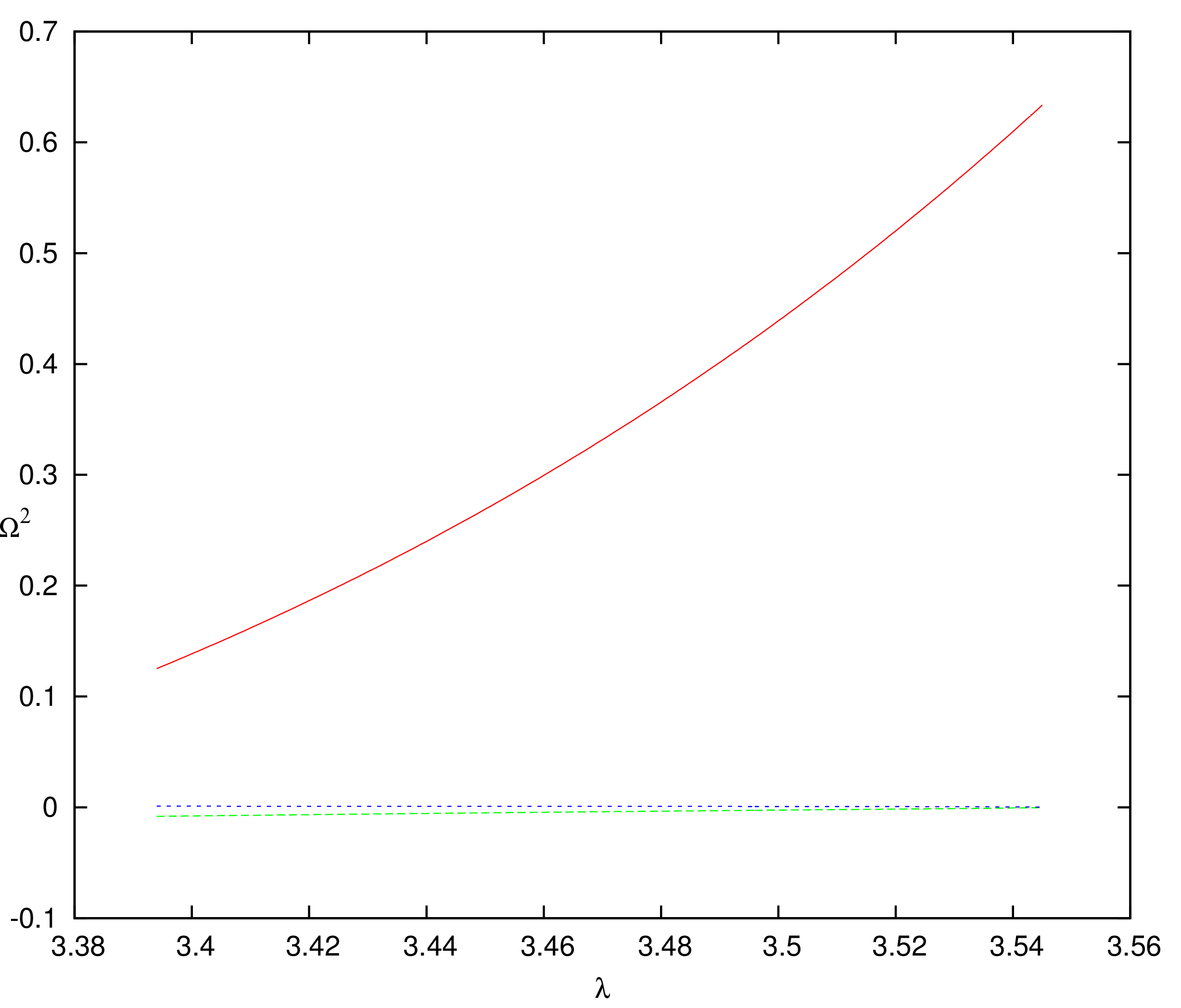}}

\caption{$\Omega^2$ projection on the $\lambda$ axis for Polytropic Case - Constant Height Model. Energy has been kept fixed at 1.06}
\end{center}
\label{some example}
\end{figure} 
 
\begin{figure}[H]
\begin{center}
\subfloat[Figure for Accretion Zone]{\includegraphics[width = 4in]{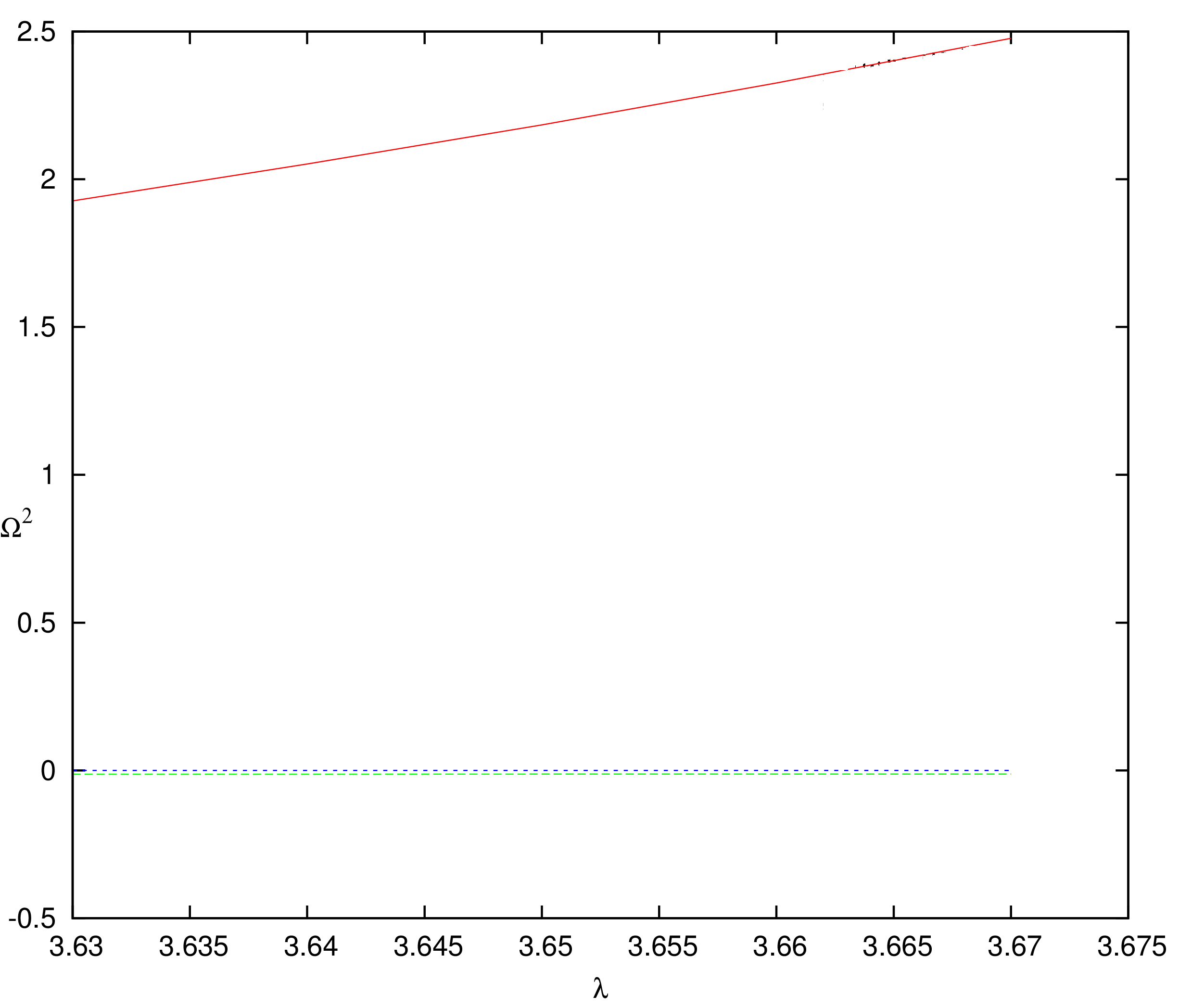}} \\
\subfloat[Figure for Wind Zone]{\includegraphics[width = 4in]{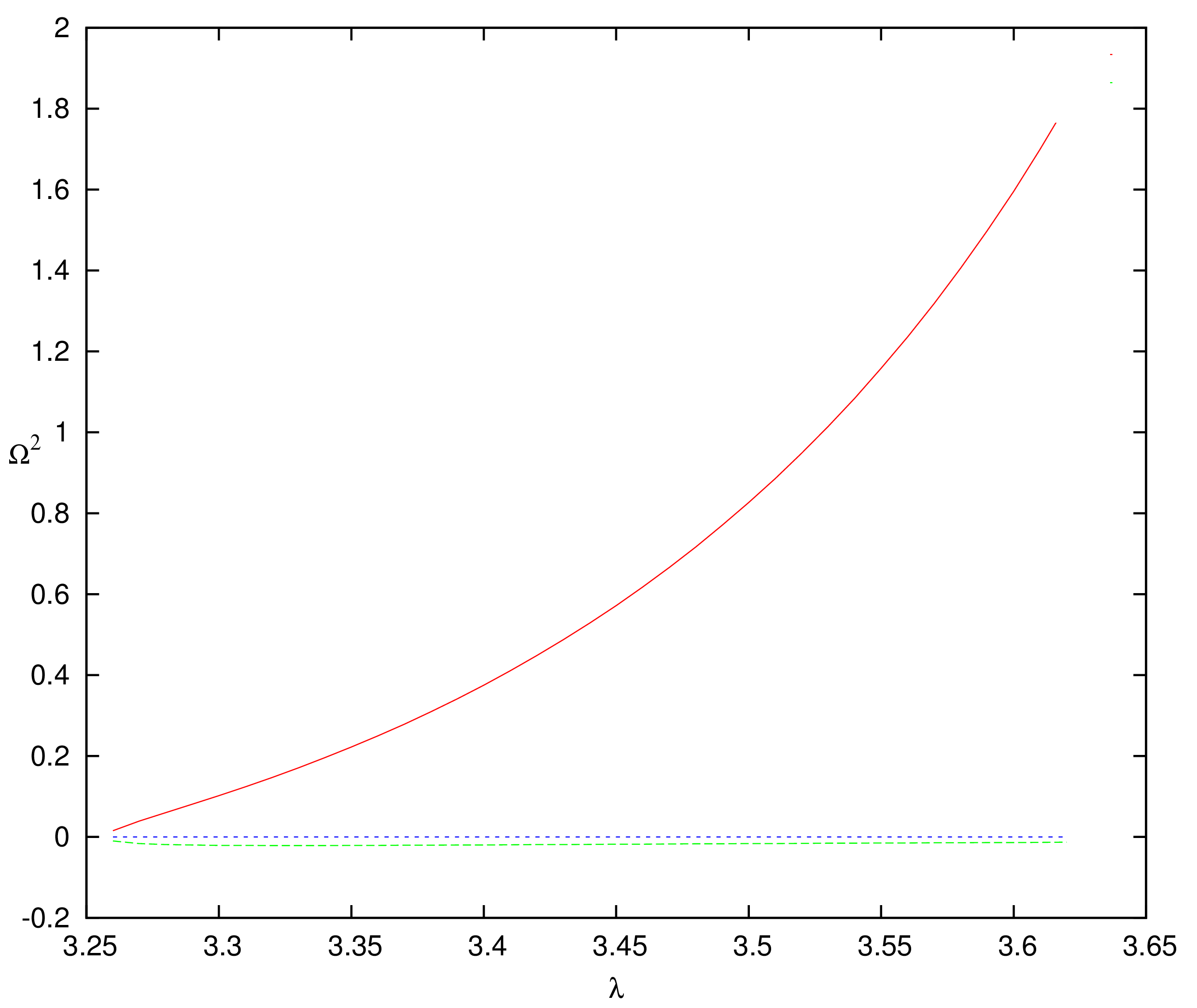}}
\end{center}
\caption{$\Omega^2$ projection on the $\lambda$ axis for Polytropic Case - Conical Equilibrium Model. Energy has been kept fixed at 1.055}
\label{some example}
\end{figure} 

\begin{figure}[H]
\begin{center}
\subfloat[Figure for Accretion Zone]{\includegraphics[width = 4in]{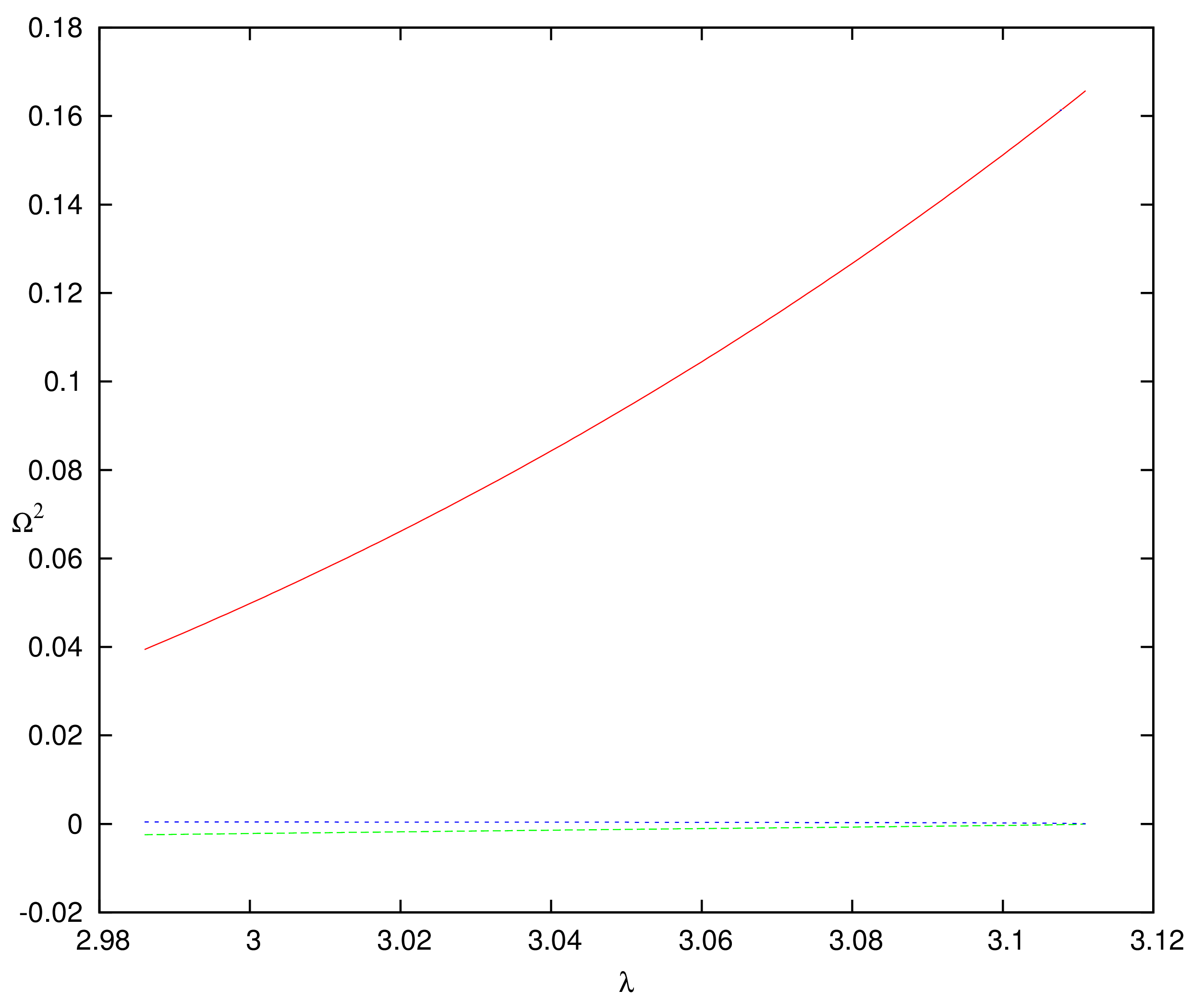}} \\
\subfloat[Figure for Wind Zone]{\includegraphics[width = 4in]{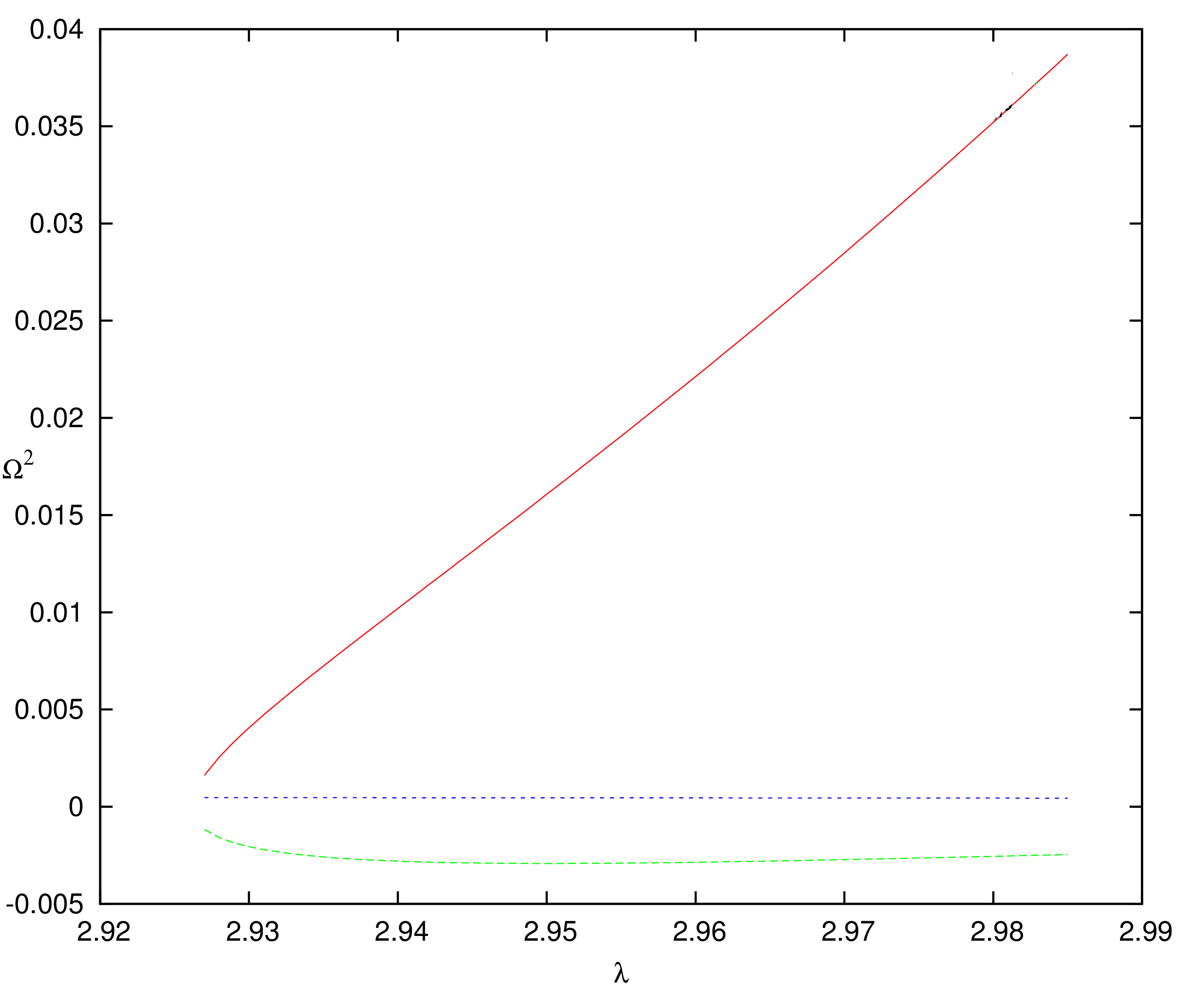}}
\end{center}
\caption{$\Omega^2$ projection on the $\lambda$ axis for Polytropic Case - Vertical Equilibrium Model. Energy has been kept fixed as 1.0055}
\label{some example}
\end{figure} 

\begin{figure}[H]
\begin{center}
\subfloat[Figure for Accretion Zone]{\includegraphics[width = 4in]{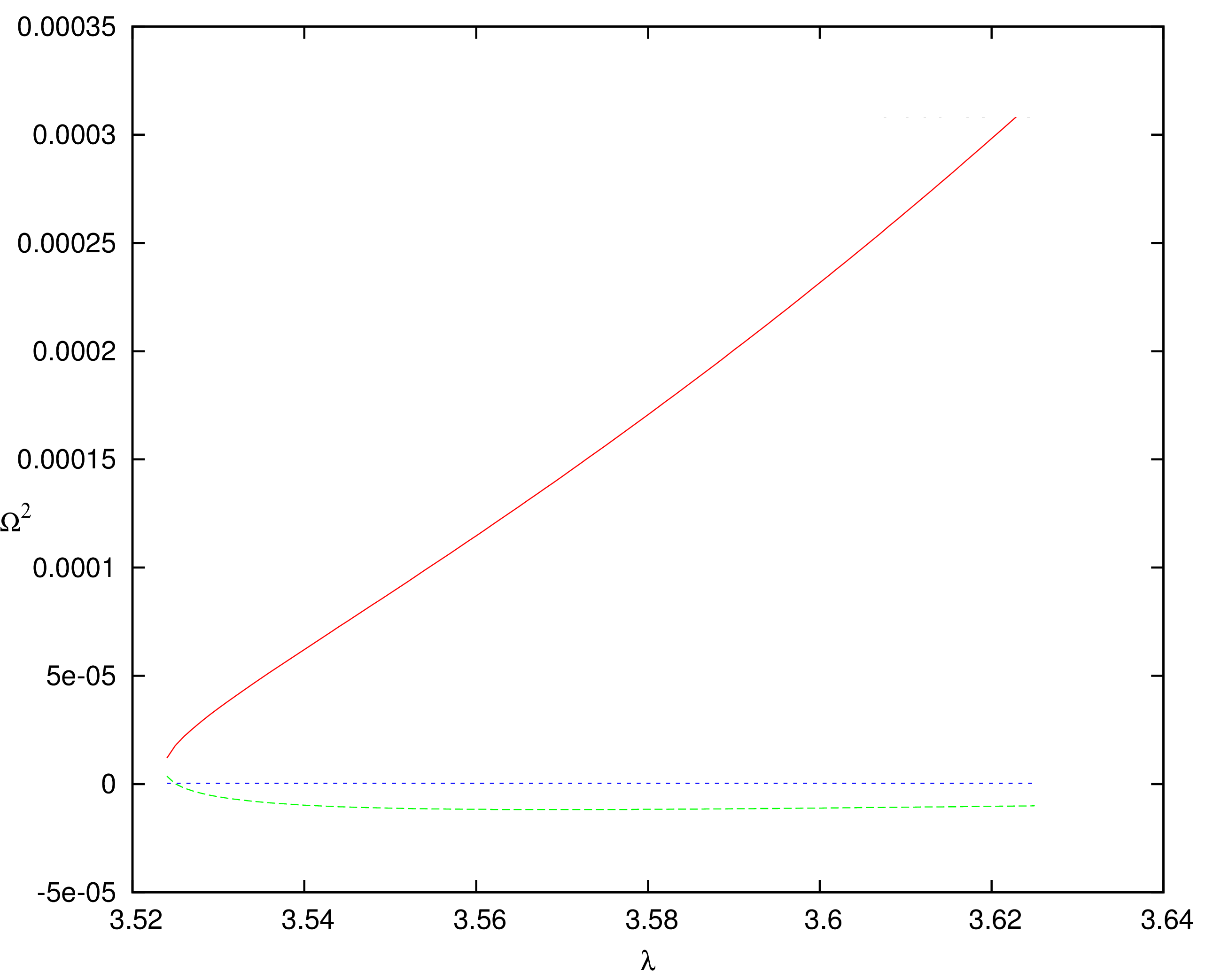}} \\
\subfloat[Figure for Wind Zone]{\includegraphics[width = 4in]{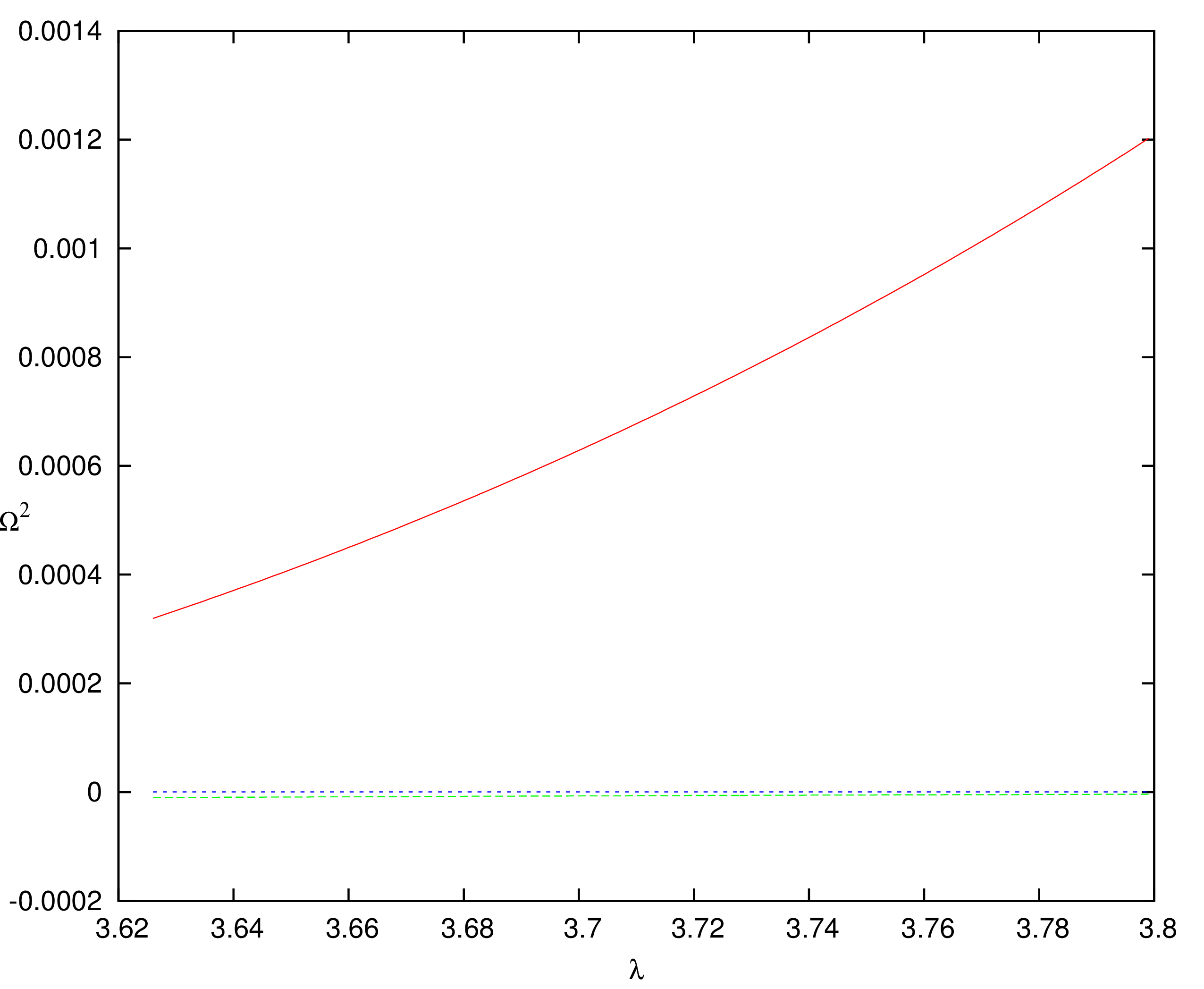}}
\end{center}
\caption{$\Omega^2$ projection on the $\lambda$ axis for Isothermal Case - Constant Height Model. Temperature has been kept fixed at $10*10^{10}K$}
\label{some example}
\end{figure} 

\begin{figure}[H]
\begin{center}
\subfloat[Figure for Accretion Zone]{\includegraphics[width = 4in]{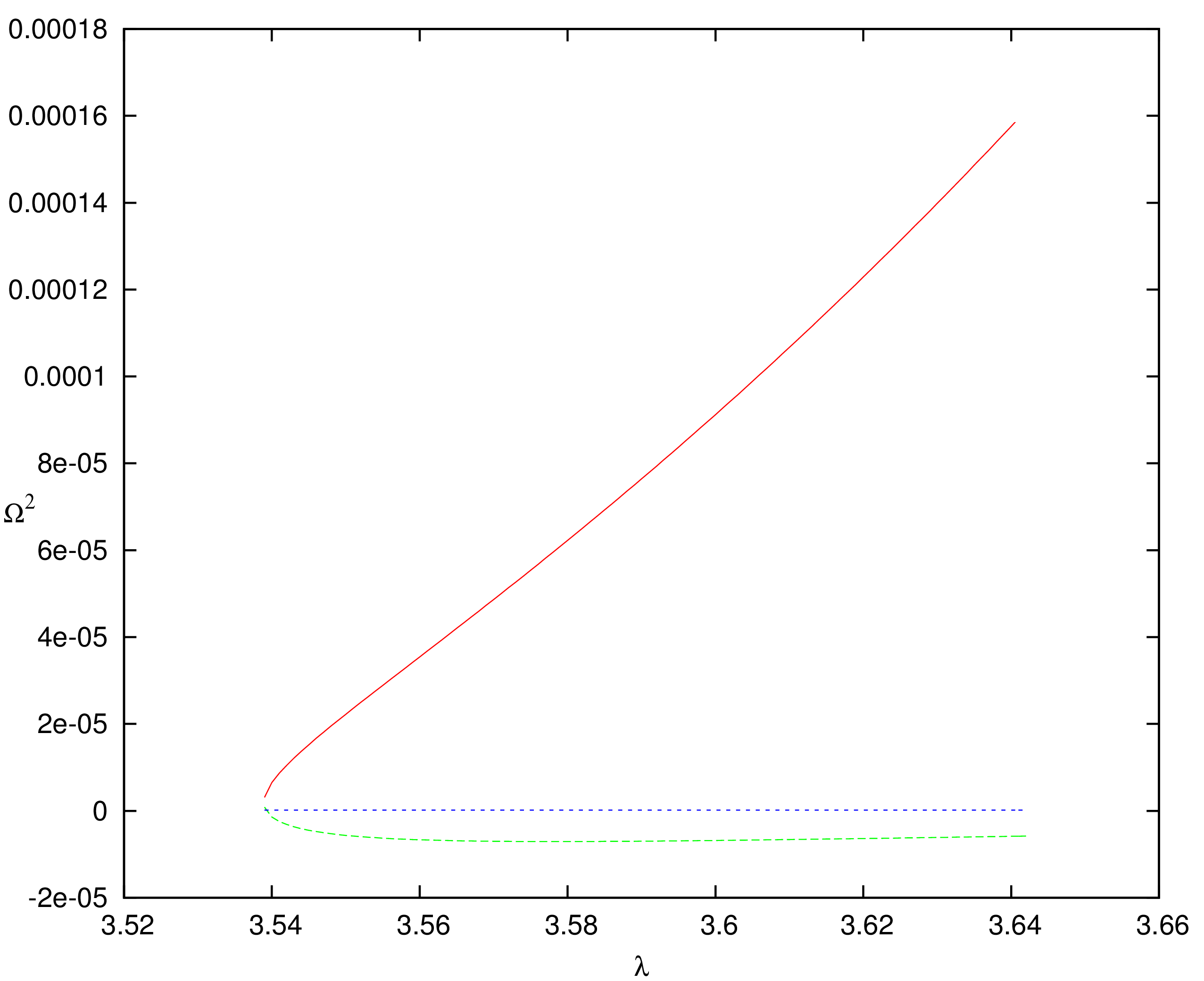}} \\
\subfloat[Figure for Wind Zone]{\includegraphics[width = 4in]{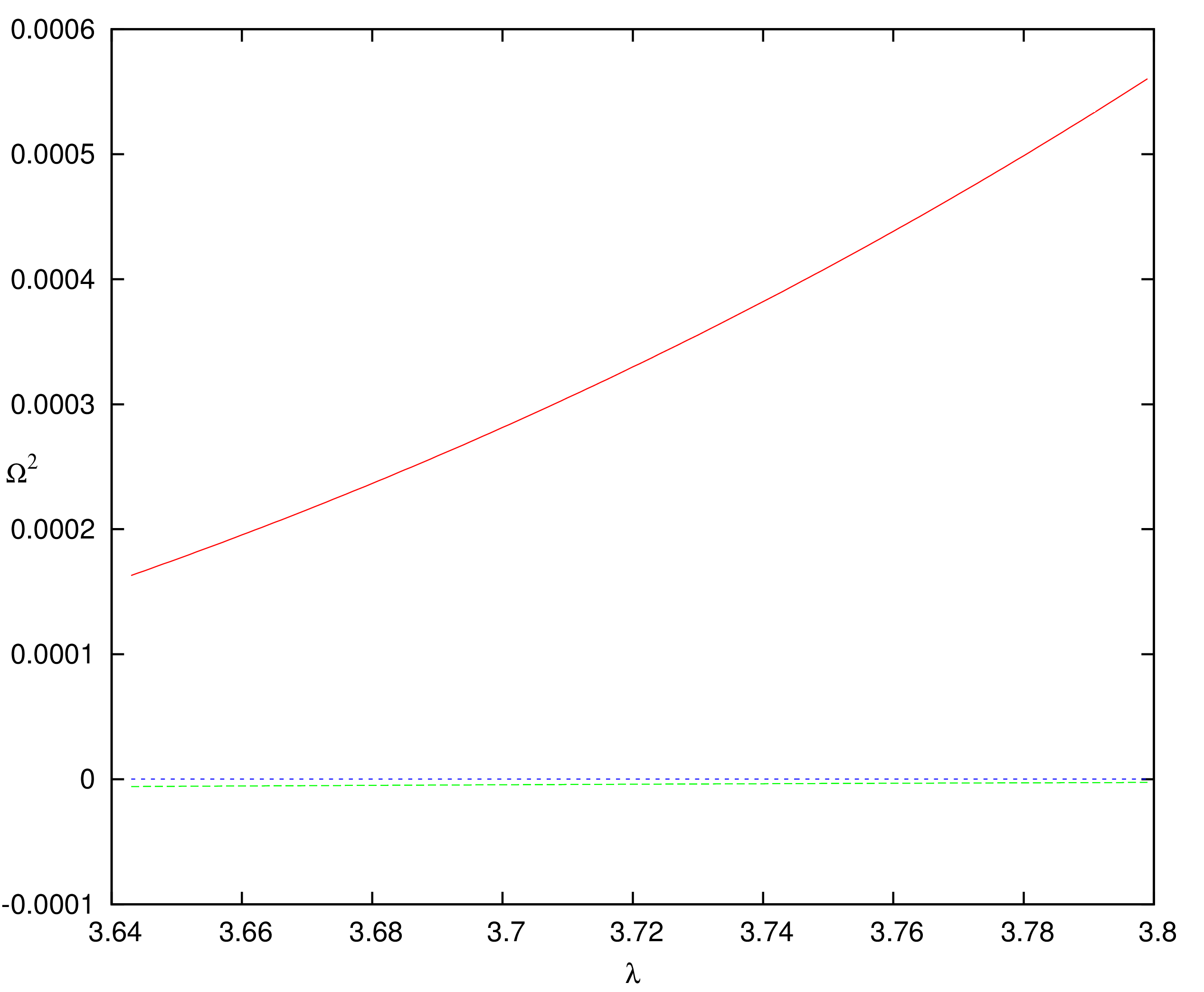}}
\end{center}
\caption{$\Omega^2$ projection on the $\lambda$ axis for Isothermal Case - Conical Equilibrium Model. Temperature has been kept fixed at $5*10^{10}K$.}
\label{some example}
\end{figure} 

\begin{figure}[H]
\begin{center}
\subfloat[Figure for Accretion Zone]{\includegraphics[width = 4in]{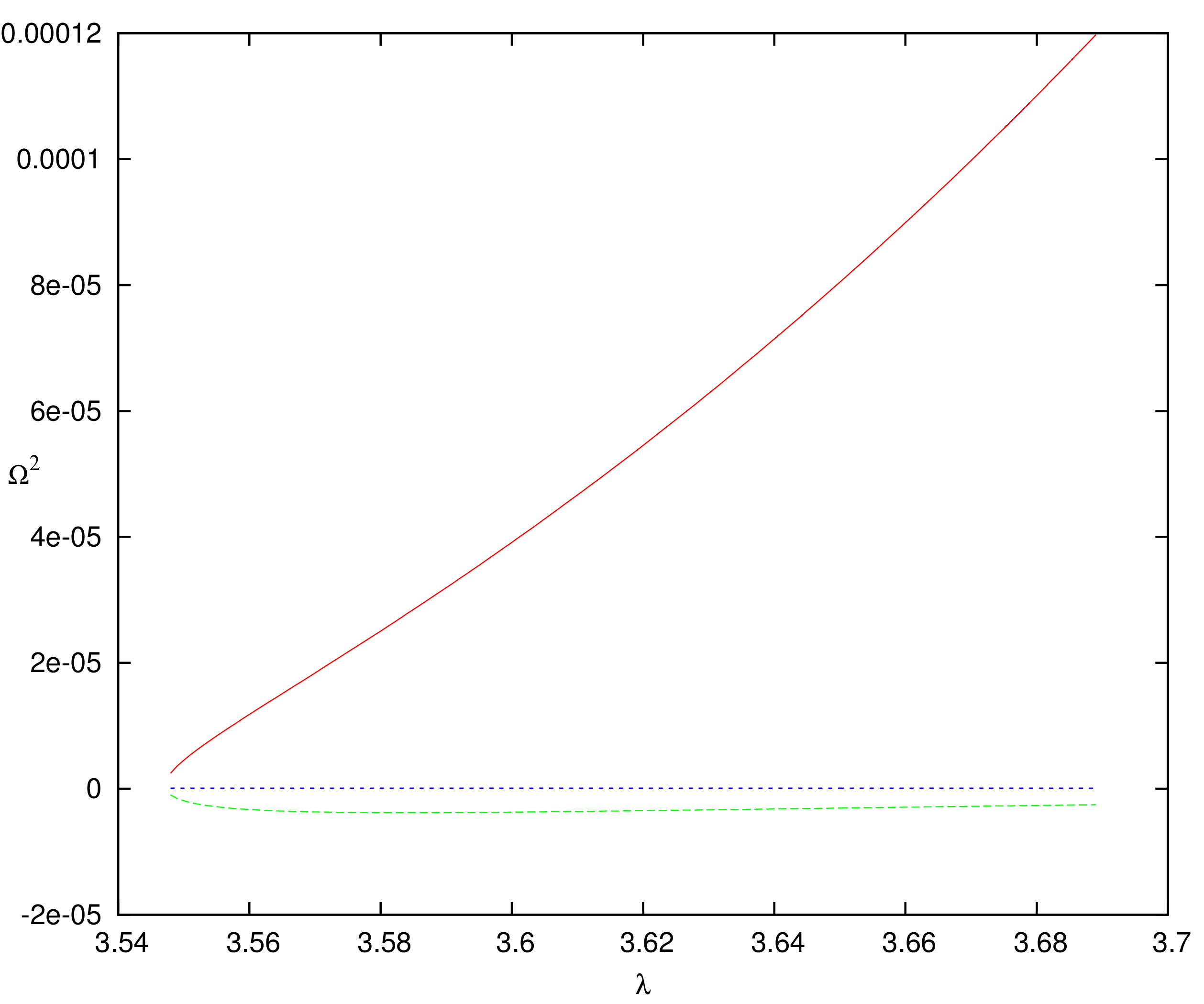}} \\
\subfloat[Figure for Wind Zone]{\includegraphics[width = 4in]{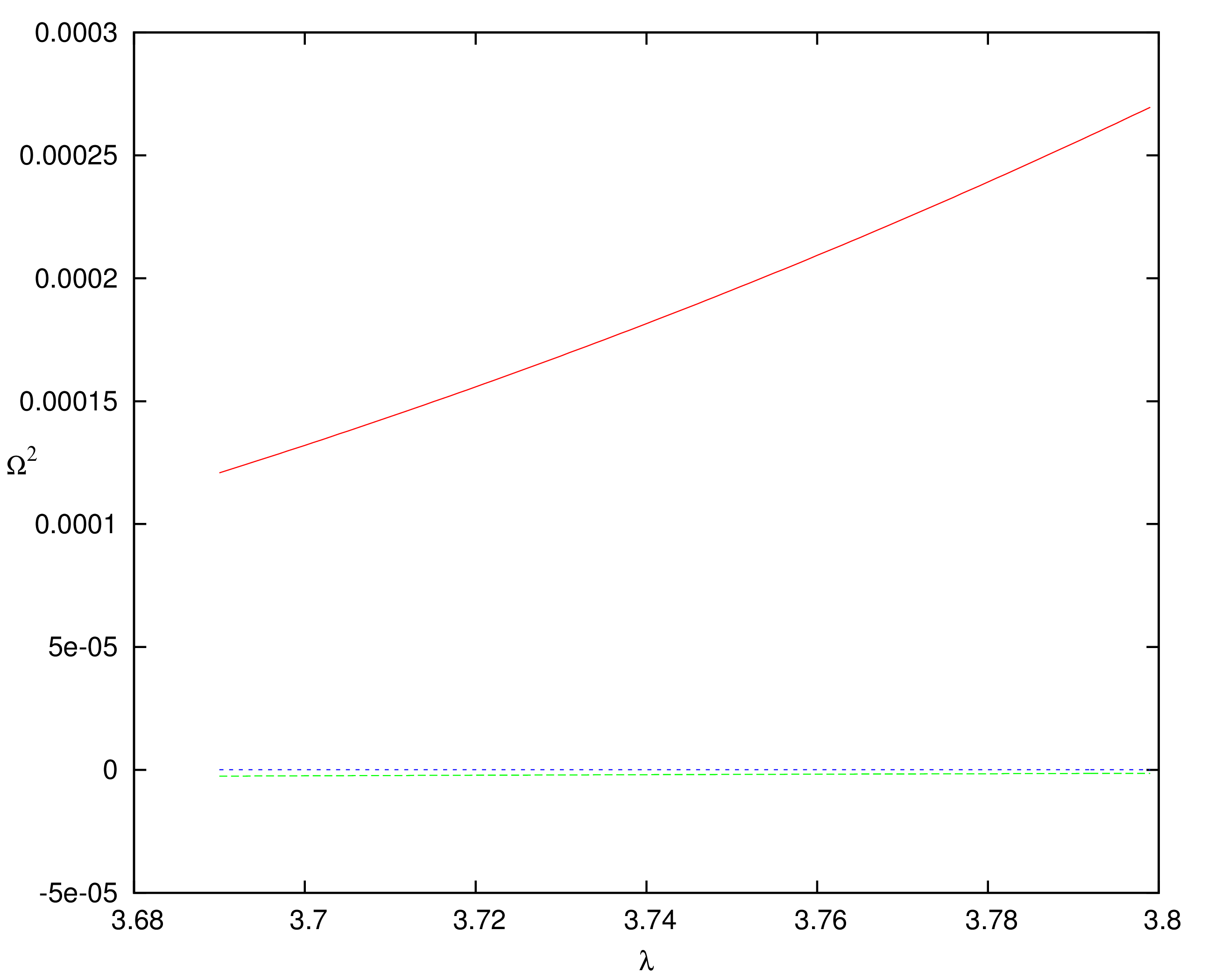}}
\end{center}
\caption{$\Omega^2$ projection on the $\lambda$ axis for Isothermal Case - Vertical Equilibrium Model. Temperature has been kept fixed at $2.5*10^{10}K$.}
\label{some example}
\end{figure}

Further, this entire treatment has been of a stationary case. Transient phenomena form the basis for evolutionary behaviour of astrophysical structures. The stability of the stationary solutions in the full general relativistic framework is still open to exploration, as well as the relevant astrophysical time scales over which they are so. This can be achieved through the usual technique of (linear) perturbation analysis of the dynamical fluid equations (governing the accretion flow) around the stationary solutions. The dispersion relation so obtained would give a comprehensive idea regarding the various time scales governing growth patterns of acoustic perturbation in the accretion disc, as well as the factors determining them. Further, it can be demonstrated that a relativistic acoustic geometry embedded by the background stationary space time may be obtained sustaining the propagation of these acoustic perturbations. This forms the basis for considering black hole systems as a classical analogue gravity model that could be naturally found in the universe. Most importantly, the system would demonstrate dual existence of event and acoustic horizons- a rich and unique, non trivial feature of this model. Acoustic surface gravity could hence be calculated, and related analogue Hawking temperature as well \cite{Tarafdar}.

\section*{Acknowledgments} 
This work was completed under Visiting Students Research Programme at HRI, Allahabad, under Prof.Tapas. K. Das (Faculty, HRI). SI would like to acknowledge all the guidance, discussions with, and suggestions of Prof.Das. The discussions with Pratik Tarafdar were also found very helpful. HRI resources that were made use of, are also duly acknowledged. 

\bibliographystyle{unsrt}

\end{document}